\documentclass[iop]{emulateapj}

\usepackage{braket}

\shorttitle{}
\shortauthors{Hasegawa et al}

\begin{document}

\title{The Origin of Heavy Element Content Trend in Giant Planets via Core Accretion}

\author{Yasuhiro Hasegawa\altaffilmark{1}, Geoffrey Bryden\altaffilmark{1}, Masahiro Ikoma\altaffilmark{2}, Gautam Vasisht\altaffilmark{1}, and Mark Swain\altaffilmark{1}}
\affil{$^1$Jet Propulsion Laboratory, California Institute of Technology, Pasadena, CA 91109, USA}
\affil{$^2$Department of Earth and Planetary Science, The University of Tokyo, 7-3-1 Hongo, Bunkyo-ku, Tokyo 113-0033, Japan}

\email{yasuhiro.hasegawa@jpl.nasa.gov}

\begin{abstract}
We explore the origin of the trend of heavy elements in observed massive exoplanets.
Coupling of better measurements of mass ($M_p$) and radius of exoplanets with planet structure models enables estimating the total heavy element mass ($M_Z$) in these planets.
The corresponding relation is characterized by a power-law profile, $M_Z \propto M_p^{3/5}$.
We develop a simplified, but physically motivated analysis to investigate how the power-law profile can be produced under the current picture of planet formation.
Making use of the existing semi-analytical formulae of accretion rates of pebbles and planetesimals,
our analysis shows that the relation can be reproduced well if it traces the final stage of planet formation.
In the stage, planets accrete solids from gapped planetesimal disks and gas accretion is limited by disk evolution.
We also find that dust accretion accompanying with gas accretion does not contribute to $M_Z$ for planets with $M_p < 10^3 M_{\oplus}$.
Our findings are broadly consistent with that of previous studies,
yet we explicitly demonstrate how planetesimal dynamics is crucial for better understanding the relation.
While our approach is simple, 
we can also reproduce the trend of a correlation between planet metallicity and $M_p$ that is obtained by detailed population synthesis calculations,
when the same assumption is adopted.
Our analysis suggests that pebble accretion would not play a direct role at the final stage of planet formation,
whereas radial drift of pebbles might be important indirectly for metal enrichment of planets.
Detailed numerical simulations and more observational data are required for confirming our analysis.
\end{abstract}


\section{Introduction} \label{sec:intro}

The detection of a large amount ($>3000$) of confirmed exoplanets has rapidly filled out a greater area in the mass-semimajor axis diagram \citep[e.g.,][]{mml11,bkb11,mls14,tjs16}.
These observations unveil a huge diversity of exoplanetary systems
that gives a number of challenges to the current theory of planet formation.
These include the presence of hot Jupiters that were first discovered by the radial velocity technique \citep{mq95}, 
the rich population of close-in super-Earths that is confirmed by both doppler and transit methods \citep[e.g.,][]{mml11,hmj10},
the existence of distant giant planetary systems that is revealed by direct imaging \citep[e.g.,][]{mzk10},
and the prediction of a significant population of free-floating planets made by microlensing observations \citep[e.g.,][]{s11,mus17}. 

A number of improvements have been made so far 
for better understanding observed exoplanetary systems and eventually developing a complete picture of planet formation.
One of the biggest leaps achieved was planetary migration \citep[e.g.,][]{gt80}.
This process arises from gravitational, tidal resonant interaction between planets and their gas disks,
and was initially invoked for explaining the presence of hot Jupiters \citep{lbr96}.
However, subsequent studies showed that migration is inevitable for planets in a wide mass range ($M_p >1 M_{\oplus}$),
and that the migration rate is generally much faster than the growth rate of planets \citep[e.g.,][]{ward97,npmk00,m01,ttw02,pbck09,hp10c}.
As a result, whereas some mechanisms for slowing down or even stopping migration have been proposed \citep[e.g.,][]{mmcf06,hp11,kl12,dmk14},
the fundamental role of planetary migration is still unclear \citep[see][for a review]{kn12}.
The general consensus in the community is that planet-forming materials move through protoplanetary disks,
and hence planet formation is a global process involved with the entire region of the disks, rather than a local process.

Characterization of exoplanets is crucial for making further progress.
For instance, influence of the host stellar metallicity on the occurrence rate of planets has been explored to specify the formation mechanism of observed exoplanets \citep[e.g.,][]{sim04,fv05,bbl14}.
Mass measurements by the radial velocity coupled with radius measurements by transit allow one to estimate the bulk density of exoplanets \citep[e.g.,][]{wm14,gcd16,jfr16}.
More recently, observations of exoplanets' atmospheres have become feasible, and one can now detect some molecules in the atmospheres \citep[e.g.,][]{tvl07,svt08,mhs11,kbd14,wsd18}.
Accompanying such observations, theoretical studies have been undertaken 
for making a link with the observations and obtaining insights into the formation and migration histories of planets \citep[e.g.,][]{il04ii,mab12,mak14,hp14,mvm16,mbj17}.
For example, \citet{gsp06} directly computed the total heavy element mass in planets, using mass and radius measurements of observed hot Jupiters \citep[also see][]{mf11}.

\begin{figure*}
\begin{minipage}{17cm}
\begin{center}
\includegraphics[width=8cm]{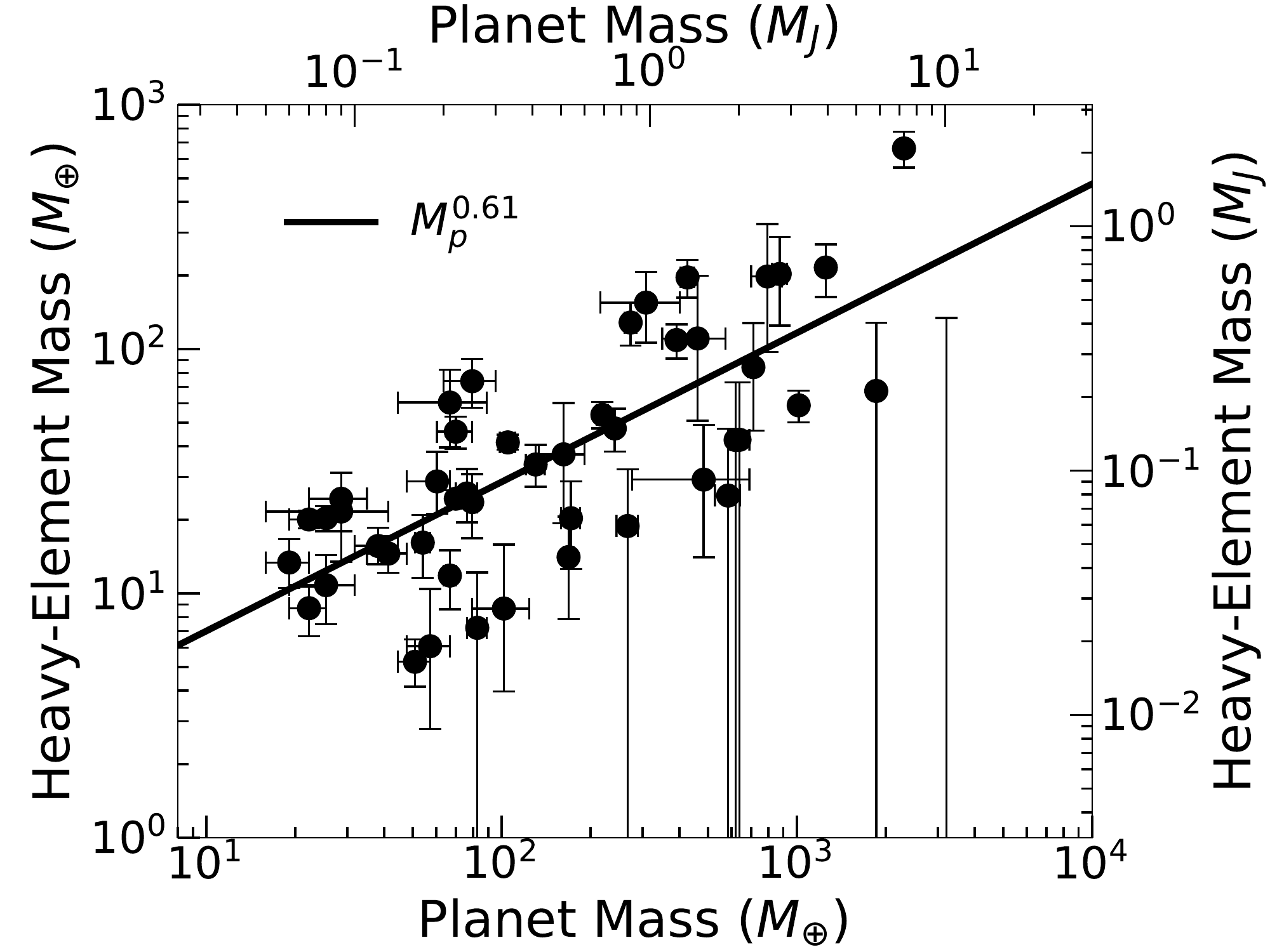}
\includegraphics[width=8cm]{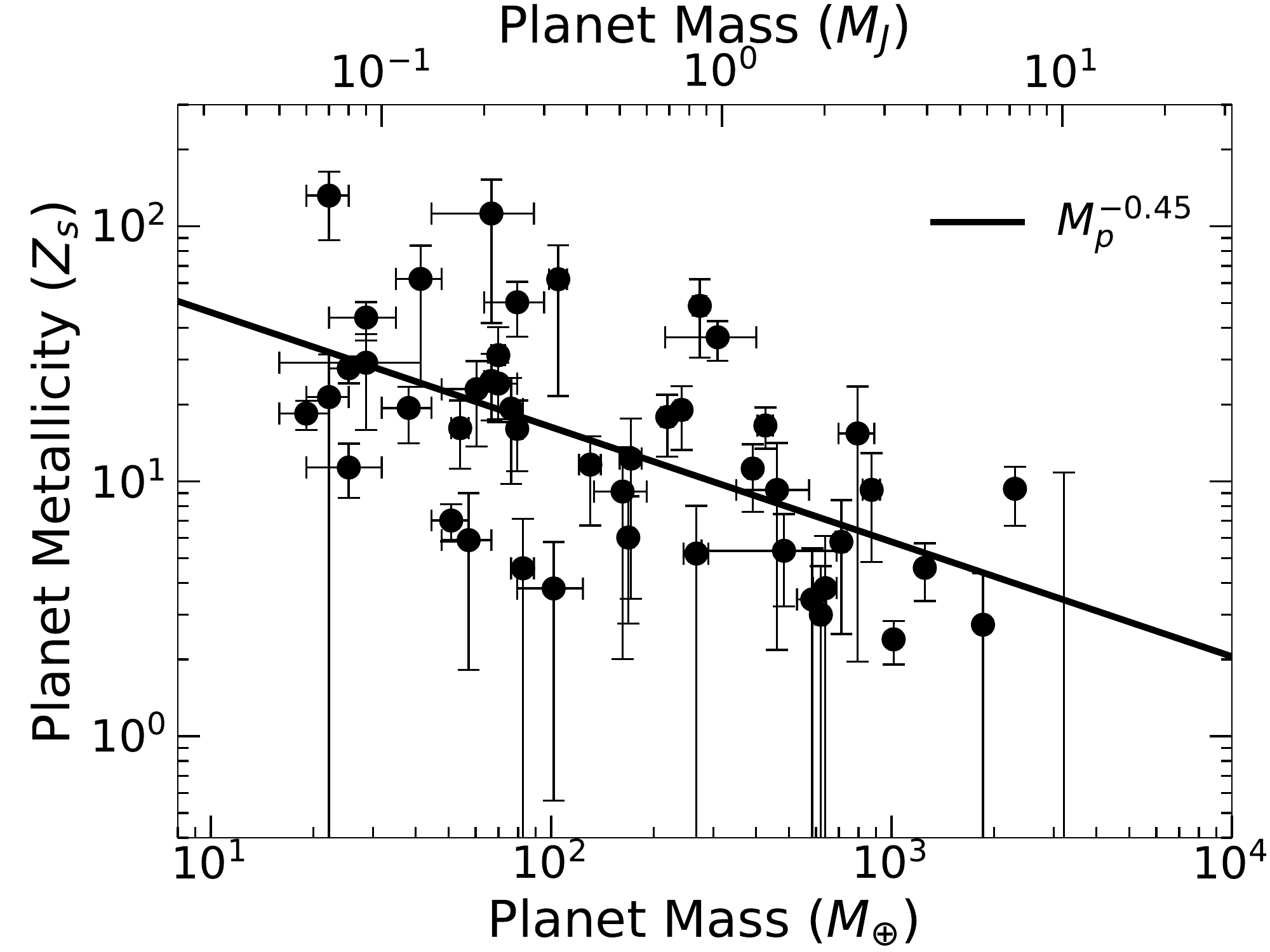}
\caption{
Reproduction of the figures made by T16.
The left panel shows the total heavy element mass ($M_Z$) as a function of planet mass ($M_p$). 
The black dots are the computed values adopted from T16 (see their Table 1), and the black solid line is their best fit (see their Figure 7).
For the right panel, the planet metallicity ($Z_p = M_Z/M_p$) that is normalized by the stellar metallicity ($Z_s$) is shown as a function of $M_p$.
As in the left panel, the black points and the black solid line are adopted from T16.
}
\label{fig1}
\end{center}
\end{minipage}
\end{figure*}

In this paper, we develop a consistent view of how accretion of gas and solids takes place onto growing planets in protoplanetary disks.
We focus on the total heavy element mass ($M_Z$) in observed exoplanets that is calculated by \citet[][hereafter T16]{tfm16}.
In their study, the radius evolution of warm Jupiters is computed, utilizing their thermal evolution models of planets (see Section \ref{sec:data_1} for the detail).
By comparing their computed planet radii with observed ones,
they specify the value of $M_Z$ for warm Jupiters and derive correlations between $M_Z$ and $M_p$ and between $M_p$ and the planet metallicity ($Z_p=M_Z/M_p$, see Figure \ref{fig1}).
Hereafter these two correlations are referred to as the $M_Z-M_p$ and the $Z_p-M_p$ relations.
In this work, we examine both planetesimal and pebble accretion within a single framework to account for the results of T16. 
More specifically, we make use of the existing semi-analytical formulae for the accretion rates of planetesimals and pebbles,
and compute the power-law indices for the $M_Z-M_p$ and $Z_p-M_p$ relations.
As clearly demonstrated below, 
we find that the subsequent planetesimal accretion after core formation is the most plausible case for better reproducing the relations.
This is consistent with the results of previous studies \citep[e.g.,][]{p96,mka14,mvm16}.
Yet, our follow-up work pins down the importance of planetesimal dynamics on the $M_Z-M_p$ relation.

The plan of this paper is as follows.
In Section \ref{sec:mod}, we describe the core accretion scenario and summarize some key quantities and equations.
In Section \ref{sec:analysis}, We develop a framework 
to investigate how both gas and solid accretion onto growing planets determine the power-law indices of the $M_Z-M_p$ and $Z_p-M_p$ relations in the core accretion picture.
We treat core formation, planetesimal accretion, pebble accretion, and the effect of gas accretion separately, 
and examine their contributions to these two relations.
In Section \ref{sec:data}, we introduce the results of T16 and reanalyze them.
We also compare the results of our theoretical analysis with those of T16.
In Section  \ref{sec:disc}, we summarize the limitation of our analysis.
We also discuss other physical processes that are not included in our analysis, 
and compare our findings with those of previous studies.
We propose a classification of observed exoplanets.
We finally list up potential roles of the current and future observations.
A brief summary and conclusions of this work are presented in Section \ref{sec:conc}.

\section{Planet formation via core accretion} \label{sec:mod}

We here consider the basic picture of core accretion.
The key quantities of this work are summarized in Table \ref{table1}.

\begin{table*}
\begin{minipage}{17cm}
\begin{center}
\caption{List of key quantities}
\label{table1}
\begin{tabular}{l|l|l} 
\hline 
Name                                                                      & Symbol                                       &  Related process                                 \\   \hline 
Host stellar metallicity                                             & $Z_s$                                        &                                                              \\ 
Total planet mass                                                   & $M_p$                                        &                                                              \\ 
Radius of planets                                                   & $R_p$                                         &                                                              \\
Total envelope mass in planets                              & $M_{XY}$                                   & Gas accretion                                        \\
Gas accretion timescales                                       & $\tau_{g,acc} (\propto M_p^D)$  &                                                                \\
Kelvin-Helmholtz timescales                                  & $\tau_{g,KH} (\propto M_p^{-d})$ & Envelope contraction ($d=4$)                \\ 
Upper limit of $\tau_{g,acc}$                                  & $\tau_{g,hydro}(\propto M_p^{-d'})$ & Disk evolution ($d'=1/3$)                        \\
Total heavy element mass in planets                     & $M_{Z}$                                       & Solid accretion                                        \\
Planet metallicity                                                    & $Z_p(=M_Z/M_p)$                       &                                                                 \\ 
Heavy element mass via gas accretion                 & $M_{Z,gas}(=Z_s M_{XY})$         & Accretion of dust via gas accretion           \\
Heavy element mass due to solid accretion          & $M_{Z,solid}$                               & Accretion of pebbles and planetesimals   \\
Core mass of planets                                             & $M_{core}$                                  &  Accretion of pebbles and planetesimals   \\
Heavy element mass via planetesimal accretion   & $M_{pl}$                                      &                                                                    \\
Heavy element mass via pebble accretion             & $M_{pe}$                                    &                                                                   \\
\hline 
\end{tabular}
\end{center}
\end{minipage}
\end{table*}

\subsection{Core formation \& gas accretion} \label{sec:mod_1}

The core accretion scenario is the widely accepted picture of how planets form in protoplanetary disks \citep[e.g.,][]{il04i,mab09,bia14}.

In this scenario, planetary cores form first and then gas accretion onto the cores proceeds with simultaneous accretion of non-negligible amounts of solids \citep[e.g.,][]{p96}.
Currently, two scenarios of core formation are actively investigated:
one is runaway and oligarchic growth and the other is pebble accretion. 
For the former, planetesimals are the dominant form of solids to build planetary cores, and their size is generally considered as a few hundred km \citep[e.g.,][]{ws89,ki98}.
In this scenario, core formation is terminated when cores accrete all the planetesimals in their feeding zone and achieve the so-called isolation mass
that is a function only of the solid surface density.
For the latter, pebble-sized ($\sim$ cm-m) particles that are weakly coupled with the disk gas provide the main contribution to core formation 
through the radial drift of such particles \citep[e.g.,][]{ok10,lj12}.
In this case, mass growth of planetary cores shuts off when the cores become massive enough to open up a gap in their gas disks \citep[e.g.,][]{ljm14,bmj18}.\footnote{
More recently, \citet{bvo17} have investigated direct core growth via pebble accretion.
Through the calculations of envelope structures around planetary cores,
they have found that the maximum mass of rocky cores that can form directly via pebble accretion is only up to $0.6 M_{\oplus}$.
They have also shown that this value is relatively insensitive to the position of the cores.
Such a small core mass arises from ablation of pebbles in planetary envelopes that prevents pebbles from reaching planetary cores.}
In other words, the cores are not exposed to the pebble flux anymore due to blocking out of pebbles by a gas gap formed around the cores.
Both the scenarios therefore lead to the final core mass that is a function only of disk parameters.

One of the key quantities in core accretion is the critical core mass that regulates the onset of efficient gas accretion onto planetary cores \citep[e.g.,][]{m80,bp86,ine00}.
The critical core mass is defined such that gaseous envelopes around the cores cannot maintain a hydrostatic equilibrium and runaway gas accretion takes place.
Under the assumption that the grain opacity of the envelopes is comparable to the ISM value,
the canonical value of $\sim 10 M_{\oplus}$ has been widely adopted in the literature \citep[e.g.,][]{p96,ine00,il04i,mab09}.
Recent studies, however, show that when dust grain growth in planetary envelopes is properly taken into account,
the value of the critical core mass tends to decrease considerably.
This arises from a lower value of the grain opacity in planetary envelopes,
which leads to rapid cooling of the envelopes and their resulting, efficient contraction \citep[e.g.,][]{mp08,hi10,mbp10,o14}.
It is interesting that a lower value ($\la 5-10M_{\oplus}$) of the critical core mass is in favor of 
theoretically reproducing the trends of observed exoplanet population \citep[e.g.,][]{mka14,hp14}.
This can be readily seen by considering gas accretion onto planetary cores (see below).
Another interesting feature of the critical core mass is that it may be used as one of the tracers to differentiate the origin of super-Earths from that of gas giants.
Given that one clear difference between these two types of planets is the envelope mass 
and that the formation mechanism(s) of super-Earths is still unclear \citep[e.g.,][]{hm13,cl13,h16},
it is of fundamental importance to identify the value of the critical core mass using the observational data of exoplanets.

Gas accretion onto planets begins once planetary cores become massive enough.
In principle, the gas accretion process can be modeled as the Kelvin-Helmholtz timescale ($\tau_{g,KH}$).
This timescale is written as \citep[e.g.,][]{ine00,il04i,hp12}
\begin{equation}
\label{eq:tau_KH}
\tau_{g,KH} = 10^{c} f_{grain} \left( \frac{M_p}{ 10 M_{\oplus} } \right)^{-d} \mbox{yr},
\end{equation}
where $f_{grain} \ll 1$ is the acceleration factor due to the reduction of grain opacity in planetary envelopes, resulting from grain growth there.
In this paper, we adopt that $c=7$ and $d=4$, following \citet[see their equation (26)]{tn97}.
As clearly seen in equation (\ref{eq:tau_KH}), $\tau_{g,KH}$ becomes much shorter than the typical disk lifetime of a few $10^6$ yrs \citep[e.g.,][]{wc11}
when the initial core mass exceeds $\sim 10 M_{\oplus}$.
This is one of the reasons why smaller core masses are preferred for reproducing the observed population of exoplanets.

One would notice that $\tau_{g,KH}$ keeps decreasing as $M_p$ increases (see equation (\ref{eq:tau_KH})). 
This can eventually lead to an unrealistically high value of the gas accretion rate ($dM_{XY}/dt$) for massive planets ($\ga 100 M_{\oplus}$).
Accordingly, an upper limit is generally imposed for limiting $dM_{XY}/dt$.
In this paper, we adopt the results of \citet{tw02}.
In their work, 2D hydrodynamical simulations are performed,
and gas accretion flow onto planets from protoplanetary disks and the fine structure of circumplanetary disks 
are resolved with high spatial resolution simulations.
They find that the upper limit of the gas accretion rate is given as \citep[see their equation (20)]{tw02}
\begin{equation}
\label{eq:tau_hydro}
\tau_{g,hydro} \simeq  1.1 \times 10^{3} \sigma_{p,acc}^{-1} \left( \frac{a_p}{5 \mbox{ au}} \right)^{1.5} \left( \frac{M_p}{10 M_{\oplus}} \right)^{-d'} \mbox{ yr},
\end{equation}
where $\sigma_{p,acc}$ is the normalized surface density of gas that participates in gas accretion, $a_p$ is the semimajor axis of planets, and $d'=1/3$.
Note that when a gap is opened up in gas disks due to disk-planet interaction \citep[e.g.,][]{npmk00,cmm06,kn12}, 
$\sigma_{p,acc}$ becomes a function of $M_p$ \citep{ti07}.

In summary, the mass growth rate of planets via gas accretion can be written as 
\begin{equation}
\label{eq:dm_xy}
\frac{dM_{XY}}{dt} \simeq \frac{dM_{p}}{dt}  = \frac{M_{p}}{\tau_{g, acc}}, 
\end{equation}
where
\begin{equation}
\label{eq:tau_gacc}
\tau_{g,acc} = \mbox{max}  \left[ \tau_{g,KH}, \tau_{g,hydro} \right].
\end{equation}
The above equations are valid mainly at the final stages of planet formation 
in which core formation nearly ends and solid accretion onto planets is insignificant, compared with gas accretion.

\subsection{Additional solid accretion} \label{sec:mod_2}

It has been suggested for a long time that additional solid accretion is essential for fully understanding the total heavy element mass of gas giant planets \citep[e.g.,][]{ppb86,ppr88}.
For instance, the enhanced metallicity in the atmosphere of Jupiter and Saturn claims the need of additional solid accretion 
during the process of forming \citep[e.g.,][]{p96,sg04}.
As another example, \citet{mka14} show that planetesimal accretion after core formation completes is important 
for reproducing the $Z_p-M_p$ relation of observed exoplanets \citep[also see][]{mvm16}.

In order to examine at what stage, how solid accretion occurs for growing planets in protoplanetary disks,
we explore the mass contribution ($M_{Z, solid}$) arising from solid accretion by decomposing it into three components:
\begin{equation}
\label{eq:mz_solid}
M_{Z, solid} =  M_{core} + M_{pl} + M_{pe}, 
\end{equation}
where $M_{core}$ is the initial, seed core mass of a protoplanet at which the subsequent gas accretion begins, 
$M_{pl}$ is the total heavy element mass that is obtained via planetesimal accretion,
and $M_{pe}$ is the total heavy element mass that is gained during accretion of small bodies such as pebbles (see Table \ref{table1}).
Accretion of both planetesimals and pebbles onto (proto)planets would be possible during the gas accretion stage \citep[e.g.,][]{p96,rr04,ambw05,tmm14}.
As described in equation (\ref{eq:mz_solid}), we treat them separately in this paper.

\subsection{Mass budget in planets} \label{sec:mod_3}

Finally, the mass budget of a planet can be written as
\begin{equation}
\label{eq:m_p}
M_p = M_{XY} + M_{Z},
\end{equation}
\begin{equation}
\label{eq:m_z}
M_Z = M_{Z, solid} + M_{Z, gas},
\end{equation}
where $M_{XY}$ is the total envelope mass of the planet, $M_Z$ is the total heavy element mass of the planet,
$M_{Z, solid}$ is the total heavy element mass that is accumulated in the planet through accretion of solids such as pebbles and planetesimals (see equation (\ref{eq:mz_solid})),
and $M_{Z, gas}$ is the total heavy element mass that is accreted through gas accretion (see Table \ref{table1}).
Note that the disk gas accreted onto planets contains small ($\sim \mu$m - mm) dust particles.
Such solids are well coupled with the disk gas and hence follow the gas motion.
Accordingly, these solids are also accumulated in planets as the gas is accreted onto the planets.
We take into account this contribution by including the term of $M_{Z,gas}$.

\section{Theoretical analysis} \label{sec:analysis}

We develop a simplified, but physically motivated analysis
to understand how accretion of gas and solids takes place onto growing protoplanets in protoplanetary disks.
We make use of the equations in the above section.

\subsection{Basic formulation} \label{sec:analysis_1}

We first formulate the basic equation exploring the $M_{Z}-M_p$ relation.

As discussed in Section \ref{sec:mod_2}, additional solid accretion would be plausible during the gas accretion stage.
It is nonetheless important to point out that the actual efficiency is currently under active investigation \citep[e.g,][]{zl07,jl17} 
and is most likely determined by disk parameters.
In order to shed light on the underlying physics, we focus only on the power index of the $M_{Z}-M_p$ relation in this paper.
While this simplification provides some limitations for our analysis (see Section \ref{sec:disc_1}),
we then need to care only about the $M_p$ dependence on each valuable.

To proceed, we adopt the approach originally developed by \citet{si08}.
In this approach, the derivative of $M_Z$ is examined, which is given as
\begin{equation}
\label{eq:dmz_dmp}
\frac{dM_Z}{dM_p} = \frac{dM_Z}{dt} \frac{dt}{dM_{p}} \approx   \frac{dM_Z}{dt} \frac{\tau_{g,acc}}{M_{p}} \propto M_p^{\Gamma'},
\end{equation}
where it is assumed that mass growth ($dM_p/dt$) of planets is dominated by gas accretion ($dM_{XY}$, also see equation  (\ref{eq:dm_xy})).
This assumption would be valid at the final stages of planet formation.

Then, we simplify the gas accretion timescale ($\tau_{g,acc}$, see equation (\ref{eq:tau_gacc})) as
\begin{equation}
\label{eq:tau_gacc_M}
\tau_{g,acc} = \mbox{max}  \left[ \tau_{g,KH}, \tau_{g,hydro} \right] \propto M_p^D,
\end{equation}
where $D=-d=-4$ when $ \tau_{g,KH}> \tau_{g,hydro}$, and $D=-d'=-1/3$ when $ \tau_{g,KH}< \tau_{g,hydro}$.
Note that we pay attention only to the $M_p$ dependence in this analysis.
Also, we neglect the effect of gas gaps that can be opened up by disk-planet interaction.
We discuss this effect in Section \ref{sec:disc_2} and Appendix \ref{app1}.

In the following, we utilize equation (\ref{eq:dmz_dmp}) and 
investigate how the power index of the $M_{Z}-M_p$ relation changes as a function of forms (planetesimals vs pebbles) of solids that are accreted onto planets.

\subsection{Contribution from planetesimal accretion} \label{sec:analysis_2}

In this section, we consider the contribution arising from $M_{pl}$ to $M_{Z,solid}$, that is, 
how planetesimal accretion proceeds in a post-stage of (initial) core formation.
Equivalently, (see equations (\ref{eq:mz_solid}), (\ref{eq:m_z}), and (\ref{eq:dmz_dmp}))
\begin{equation}
\label{eq:mz_pl}
M_{Z} \approx M_{Z, solid} \approx  M_{pl},
\end{equation}
\begin{equation}
\label{eq:dmz_dmp_pl}
\frac{dM_Z}{dM_p}  \approx   \frac{dM_{pl}}{dt} \frac{\tau_{g,acc}}{M_{p}} \propto M_p^{\Gamma'_{pl}}.
\end{equation}

The remarkable recognition that continuous accretion of planetesimals is important for planet formation is made by the milestone work of \citet{p96}.
In this study, it is assumed that such accretion originates from the expansion of planets' feeding zone as the planets grow in mass and their Hill radius increases.
Adopting the most efficient accretion rate of planetesimals, 
they can reproduce the trend of the enhanced atmospheric metallicity of Jovian planets in the solar system such as Jupiter and Saturn.
Such efficient accretion of planetesimals leads to emergence of the so-called "phase 2",
where the planetesimal accretion rate is so high ($\sim 10^{-6} M_{\oplus}$ yr$^{-1}$) that the onset of runaway gas accretion is postponed for $\sim$ a few Myr.
Despite of the success achieved by their model,
a number of follow-up studies pose a question about their assumption 
that the most efficient planetesimal accretion would be realized and continue for a long ($\sim$ Myr) time \citep[e.g.,][]{fbb07,zl07,si08,hp14}.
This is because, following mass growth of planets, planetesimals in their feeding zone will be used up,
and some of them will be even scattered out of the zone due to the gravitational interaction with the planets.
Coupled with the eccentricity dumping by the disk gas,
this scattering process can end up with the creation of a gap in planetesimal disks around planets.
In fact, the common conclusion of these studies is that when both the dynamics of planetesimals in gas disks and the effect of planetary growth are considered realistically,
efficient planetesimal accretion cannot be established.

To appropriately take into account the dynamics of planetesimals around a growing planet in a gas disk
and to reliably derive the power-law index ($\Gamma'_{pl}$) of $dM_Z/dM_p$ (see equation (\ref{eq:dmz_dmp_pl})),
we here make use of the results of \citet{si08}.
In their study, a number of $N-$body simulations are carried out to investigate how planetesimal accretion takes place for planets that undergo gas accretion,
and to derive a semi-analytical accretion rate of planetesimals ($dM_{pl}/dt$).
Based on their results, $dM_{pl}/dt$ is determined by the interplay among excitation of planetesimals' eccentricity by a growing planet, 
dumping of their eccentricity by the disk gas, and the expansion of the Hill radius of the planet.
When the dumping efficiency of planetesimals' eccentricity by the disk gas is less than the expansion rate of the Hill radius of a growing planet,
the growth rate of the planet is so fast that the planet can keep accreting planetesimals in its expanding feeding zone.
In other words, a gap is not generated in the planetesimal disk.
For this case, the planetesimal accretion rate is given as (see equations (22) and (24) in \citet{si08})
\begin{equation}
\label{eq:si08_24}
\left( \frac{dM_{pl}}{dt} \right)_{nogap}  \propto R_p^2 M_p^{-\alpha/3} \tau_{g, acc}^{-\alpha} \propto M_p^{(2-\alpha)/3} \tau_{g,acc}^{-\alpha},
\end{equation}
where $\alpha \simeq 4/5$.
Note that $dM_{pl}/dt$ is a function of $\tau_{g,acc}$.
This originates from that planetary growth is regulated mainly by gas accretion. 
On the other hand, when the eccentricity dumping of scattered planetesimals by the disk gas is more significant than the Hill radius expansion, 
then planetary growth is slow enough that planetesimals can leave from the feeding zone of a planet before they will be accreted.
Equivalently, a gap can open up in planetesimal disks.
Under this situation, 
the accretion rate of planetesimals is written as (see equations (23) and (25) in \citet{si08})
\begin{equation}
\label{eq:si08_25}
\left( \frac{dM_{pl}}{dt} \right)_{gap}  \propto R_p^2 M_p^{-\alpha' /6} \tau_{g, acc}^{-\alpha'} \propto M_p^{(4-\alpha')/6} \tau_{g,acc}^{-\alpha'}
\end{equation}
where $\alpha' \simeq 7/5$.
Again, $dM_{pl}/dt$ is related to $\tau_{g,acc}$.
Thus, the planetesimal accretion rate is a function of both $M_p$ and $\tau_{g,acc}$, 
and the functional forms of $dM_{pl}/dt$ are different, depending on the creation of a gap in planetesimal disks.

We are now in a position to derive the power-law index of $dM_{Z}/dM_p$, 
which is given as (with equation (\ref{eq:tau_gacc_M}))
\begin{equation}
\label{eq:dz_dm_pl_nogap}
\Gamma'_{pl}  =  -\frac{1 + \alpha }{3} + D ( 1- \alpha ) = \frac{ D - 3 }{5} 
\end{equation}
without planetesimal gaps, and 
\begin{equation}
\label{eq:dz_dm_pl_gap}
\Gamma'_{pl}  =  -\frac{2 + \alpha' }{6} + D ( 1- \alpha' ) = -\frac{12D + 17}{30} 
\end{equation}
with planetesimal gaps.
Given that there are two modes in gas accretion (see equation (\ref{eq:tau_gacc})), 
one of which is regulated by the Kelvin-Helmholtz timescale, the other of which is limited by disk evolution,
the corresponding power-law indices are summarized in Table \ref{table2}.
By integrating $dM_{Z}/dM_p$,
we find the resulting power-law indices of $M_{Z} (\propto M_p^{\Gamma_{pl}})$ for planetesimal accretion (see Table \ref{table3}):
\begin{equation}
\label{eq:mz_solid_pl}
\Gamma_{pl} =
                                    \left\{ \begin{tabular}{@{}l@{}} 
                                                                       $-2/5$  with no gap and $\tau_{g,acc} = \tau_{g,KH}$ \\
                                                                       $1/3$  with no gap and $\tau_{g,acc} = \tau_{g,hydro}$  \\
                                                                       $2$  with a gap and $\tau_{g,acc} = \tau_{g,KH}$ \\
                                                                       $3/5$  with a gap and $\tau_{g,acc} = \tau_{g,hydro}$.  \\
                                            \end{tabular} \right.
\end{equation}

Based on the above analysis, the power-law index of $Z_p (\propto M_p^{\beta_{pl}})$ is the same as $\Gamma'_{pl}$ and is given as (also see Table \ref{table2})
\begin{equation}
\label{eq:zp_solid_pl}
\beta_{pl} \propto
                                    \left\{ \begin{tabular}{@{}l@{}}                                     
                                                                       $-7/5$  with no gap and $\tau_{g,acc} = \tau_{g,KH}$ \\
                                                                       $-2/3$  with no gap and $\tau_{g,acc} = \tau_{g,hydro}$  \\
                                                                       $1$  with a gap and $\tau_{g,acc} = \tau_{g,KH}$ \\
                                                                       $-2/5$  with a gap and $\tau_{g,acc} = \tau_{g,hydro}$.  \\                                   
                                            \end{tabular} \right.
\end{equation}
It is interesting that our analysis predicts that $\beta_{pl} =1$ for the case with planetesimal gaps and $\tau_{g,acc} = \tau_{g,KH}$,
which is inconsistent with the current trend of observed exoplanets.
We consider that this inconsistency suggests that such a case never occurs in planet formation.
In fact, it can be expected readily that if planets are massive enough to open up a gap in planetesimal disks,
the corresponding $\tau_{g,KH}$ should be smaller than $\tau_{g,hydro}$ (see equation (\ref{eq:tau_gacc_M})).
Our case study therefore would be useful for specifying the mass growth path of planets without any detailed calculations.

Thus, we find that the $M_Z-M_p$ relation has different slopes, depending on the planetesimal distribution around planets and their gas accretion rates.

\begin{table*}
\begin{minipage}{17cm}
\begin{center}
\caption{Power-law indices of $dM_{Z}/dM_p(\propto M_p^{\Gamma'})$ for both cases of planetesimal and pebble accretion}
\label{table2}
\begin{tabular}{l|c|c|c} 
\hline 
Gas accretion mode                            & Planetesimal Accretion      & Planetesimal Accretion     & Pebble Accretion                               \\ 
                                                            & No Gap                              & Gap                                   &                                              \\  \hline 
Kelvin-Helmholtz ($D=-4$)                 & $- 7/5 $                              & $ 31/30 \simeq  1$         & -13/3                       \\ 
Limited by disk evolution ($D=-1/3$)  & $ - 2/3 $                             & $ - 13/30 \simeq - 2/5$      & -2/3   \\
\hline 
\end{tabular}
\end{center}
\end{minipage}
\end{table*}

\subsection{Contribution from pebble accretion} \label{sec:analysis_3}

We here examine the case of pebble accretion.
Equivalently, we consider the following case (see equations (\ref{eq:mz_solid}), (\ref{eq:m_z}), and (\ref{eq:dmz_dmp}):
\begin{equation}
\label{eq:mz_pl}
M_{Z} \approx M_{Z, solid} \approx  M_{pe}. 
\end{equation}
\begin{equation}
\label{eq:dmz_dmp_pe}
\frac{dM_Z}{dM_p}  \approx   \frac{dM_{pe}}{dt} \frac{\tau_{g,acc}}{M_{p}} \propto M_p^{\Gamma'_{pe}}.
\end{equation}

Substantial progress is currently being made for pebble accretion since the first realization of its importance on planet formation \citep[see][as a most recent review]{jl17}.
For the completeness of this paper, we will utilize the most recent results of pebble accretion and develop a formulation, 
which is similar to that of planetesimal accretion (see Section \ref{sec:analysis_2}).
It is nonetheless fair to mention that pebble accretion is not explored at the final stages of gas giant formation very much, compared with that of planetesimal accretion.
In fact, even in the most recent studies, the primary target is the role of pebble accretion on core formation  \citep[e.g.,][]{blj15,mbj17}.
Furthermore, these studies essentially treat accretion of gas and pebbles onto planets separately.
In other words, the adopted pebble accretion rate ($dM_{pe}/dt$) is independent of the gas accretion rate.
The following analysis, therefore, should be viewed as a reference one, rather than the final results.
Once the similar level of complexity is included in numerical simulations of pebble accretion,
one can undertake a more comprehensive calculation 
to examine the importance of pebble accretion on the $M_{Z}-M_p$ and the $Z_{p}-M_p$ relations more realistically.

Keeping this caveat in mind, 
we discuss the accretion rate of pebbles onto growing planets.
In practice, $dM_{pe}/dt$ is written as \citep[see equation (34) in][]{jl17}
\begin{equation}
\frac{dM_{pe}}{dt}  \propto M_p^{2/3},
\end{equation}
where the so-called Hill regime is considered. 
This is because our analysis assumes that (initial) core formation is almost completed and the core mass should be relatively large ($\ga 1-5 M_{\oplus}$).
For this case, the growth mode is regulated by the relative velocity of Keplerian shear, rather than the azimuthal drift \citep[e.g.,][]{ok10,lj12,igm16}.
Then, the power-law index of $dM_{pe}/dM_p (\propto M_p^{\Gamma'_{pe}})$ can be calculated as
\begin{equation}
\Gamma'_{pe} = D -\frac{1}{3} 
\label{eq:dz_dm_pe}
\end{equation}
Table \ref{table2} summarizes the results for both the cases of gas accretion ($\tau_{g,acc}=\tau_{g,KH}$ and $\tau_{g,acc}=\tau_{g,hydro}$).

When integrating the above equation, we obtain the power-law index of $M_{Z}( \propto M_p^{\Gamma_{pe}})$ for pebble accretion,
which is given as (see Table \ref{table3})
\begin{equation}
\label{eq:mz_solid_pe}
\Gamma_{pe} =
                                    \left\{ \begin{tabular}{@{}l@{}} 
                                                                       $-10/3$  with $\tau_{g,acc} = \tau_{g,KH}$ \\
                                                                       $1/3$  with $\tau_{g,acc} = \tau_{g,hydro}$.  \\
                                            \end{tabular} \right.
\end{equation}
Also, the power-law index of $Z_p (\propto M_p^{\beta_{pe}})$ is written as
\begin{equation}
\label{eq:zp_solid_pe}
\beta_{pe} =
                                    \left\{ \begin{tabular}{@{}l@{}} 
                                                                       $-13/3$  with $\tau_{g,acc} = \tau_{g,KH}$ \\
                                                                       $-2/3$  with $\tau_{g,acc} = \tau_{g,hydro}$.  \\
                                            \end{tabular} \right.
\end{equation}

As in the case with planetesimal accretion, the $M_Z-M_p$ relation has different slopes for different gas accretion recipes.

\subsection{Contribution from planetary cores} \label{sec:analysis_4}

In this section, we focus on the contribution of $M_{Z,soild}$ arising from core formation (see equation (\ref{eq:mz_solid})): 
\begin{equation}
\label{eq:mz_core}
M_{Z} \approx M_{Z, solid} \approx  M_{core}. 
\end{equation}

As discussed in Section \ref{sec:mod_1}, both the oligarchic growth and pebble accretion scenarios lead to the core mass that is independent of $M_p$.
Then the power-law indices of $M_Z(\propto M_p^{\Gamma_{core}})$ and $Z_p(\propto M_p^{\beta_{core}})$ are readily computed as
\begin{equation}
\label{eq:mz_p_core}
\Gamma_{core} = \mbox{constant},
\end{equation}
\begin{equation}
\label{eq:z_p_core}
\beta_{core} = -1.
\end{equation}
It is interesting that these profiles are inconsistent with the trend of observed exoplanets (see Figure \ref{fig1}, also see Section \ref{sec:data}).

\subsection{Contribution arising from gas accretion} \label{sec:analysis_5}

Finally, we examine the contribution ($M_{Z,gas}$) originating from gas accretion (see equation (\ref{eq:m_z})). 

For this case, we can directly compute the total amount of $M_{Z,gas}$.
Assuming that the dust abundance in the gas accreted onto planets is comparable to $Z_s$,
the value of $M_{Z,gas}$ can be given as (using equation (\ref{eq:m_p}))
\begin{equation}
\label{eq:mz_gas_mz_solid}
M_{Z,gas}  \equiv Z_s M_{XY} =  Z_s(M_p - M_Z).
\end{equation}
Given that $M_p \gg M_Z$ for gas giant planets, the contribution of $M_{Z,gas}$ is only about 1 \% ($\sim Z_s$) of the total planet mass.
We thus can conclude that dust accretion accompanying with gas accretion is not significant to $M_Z$ for planets with the mass of  $M_p \ga 10^3 M_{\oplus}$.
As shown below (see Section \ref{sec:data_4}), this conclusion is justified for observed massive exoplanets.

\section{Reanalysis of the results of T16} \label{sec:data}

We here turn our attention to the results obtained by T16.
We reanalyze their computed values of the total heavy element mass in observed exoplanets and investigate how they are useful for developing a better understanding of planet formation.

\subsection{The results of T16} \label{sec:data_1}

We first introduce the results of T16 \citep[see Figure \ref{fig1}, also see][]{mf11}.

In the study, observed exoplanets that have better measurements of mass and radius 
are chosen from the Extrasolar Planets Encyclopedia \citep[exoplanets.eu][]{sdl11} and the NASA Exoplanet Archive \citep{acc13}.
Especially, 47 exoplanets are selected from larger samples based on the criterion of a relatively low value of stellar insolation ($F_* < 2 \times 10^8$ erg s$^{-1}$ cm$^{-2}$).
This criterion is adopted in order to filter out potentially inflated hot Jupiters, the origin of which is still unknown.

Through the careful examination of the data from both the original sources and the websites,
they obtain the values of the planet mass ($M_p$) and radius ($R_p$), and the host star age and metallicity ($Z_s$).
They make use of these values to combine their planet structure model and to compute the thermal evolution of planets.
Such computations allow one to trace the radius evolution of planets.
More specifically, they adopt 1D planet structure models that are composed of an inert core (a 50/50 rock-ice mixture), 
homogenous convective envelope (a H/He-rocl-ice mixture), and a radiative atmosphere as the upper boundary condition.
For the atmosphere model, the solar metallicity grids are interpolated from \citet{fnb07}.
Their calculations employ a number of assumptions and simplifications.
A more detailed model should include a self-consistent treatment of atmospheres, the composition of heavy elements, the treatment of thermal properties of cores (see section 3 of T16).
They however find that uncertainties from observations (mass, radius, and host star age) are still dominant over those from model uncertainties (see Section \ref{sec:disc_1}).
By comparing the computed radius of planets with the observational data,
they identify the values of $M_Z$ in the planets that can distribute in both their cores and envelopes.

Here we simply summarize their derived $M_Z-M_p$ and $Z_p-M_P$ relations (also see their Figures (7) and (11)):
\begin{equation}
\label{eq:m_z_obs}
M_Z \propto M_p^{\Gamma_{T16}},
\end{equation}
\begin{equation}
\label{eq:z_p_obs}
\frac{Z_p}{Z_s} = \frac{M_Z}{M_p} \frac{1}{Z_s} \propto M_p^{\beta_{T16}},
\end{equation}
where $\Gamma_{T16}=0.61 \pm 0.08 $ and $\beta_{T16}=-0.45 \pm 0.09$.
In this paper, we adopt that $\Gamma_{T16} \approx 3/5$ and $\beta_{T16} \approx -2/5$, respectively.
For clear presentation, we do not show error bars in figures in this and following sections.
It is interesting that $\beta_{T16} \approx \Gamma_{T16} -1$.
This suggests that both $M_Z$ and $M_p$ are almost independent of or only very weakly dependent on $Z_s$ for observed exoplanets.
In fact, exoplanet observations confirm that while the occurrence rate of exoplanets is correlated with stellar metallicity \citep[e.g.,][]{fv05,bbl14,hp14},
the maximum mass of planets is not related to $Z_s$.
Note that T16 found that the $Z_p-M_p$ relation becomes clearer when the planet metallicity is normalized by the host stellar metallicity (see their figures 10 and 11).
Accordingly, we adopt the same convention.

In the following, we reanalyze the results of T16 in order to derive some constraints on planet formation and to examine how the $M_Z-M_p$ relation can be reproduced.

\subsection{The envelope mass and the critical core mass} \label{sec:data_2}

We begin with computing the envelope mass ($M_{XY}$, see equation (\ref{eq:m_p})) and considering the critical core mass.

\begin{figure*}
\begin{minipage}{17cm}
\begin{center}
\includegraphics[width=8cm]{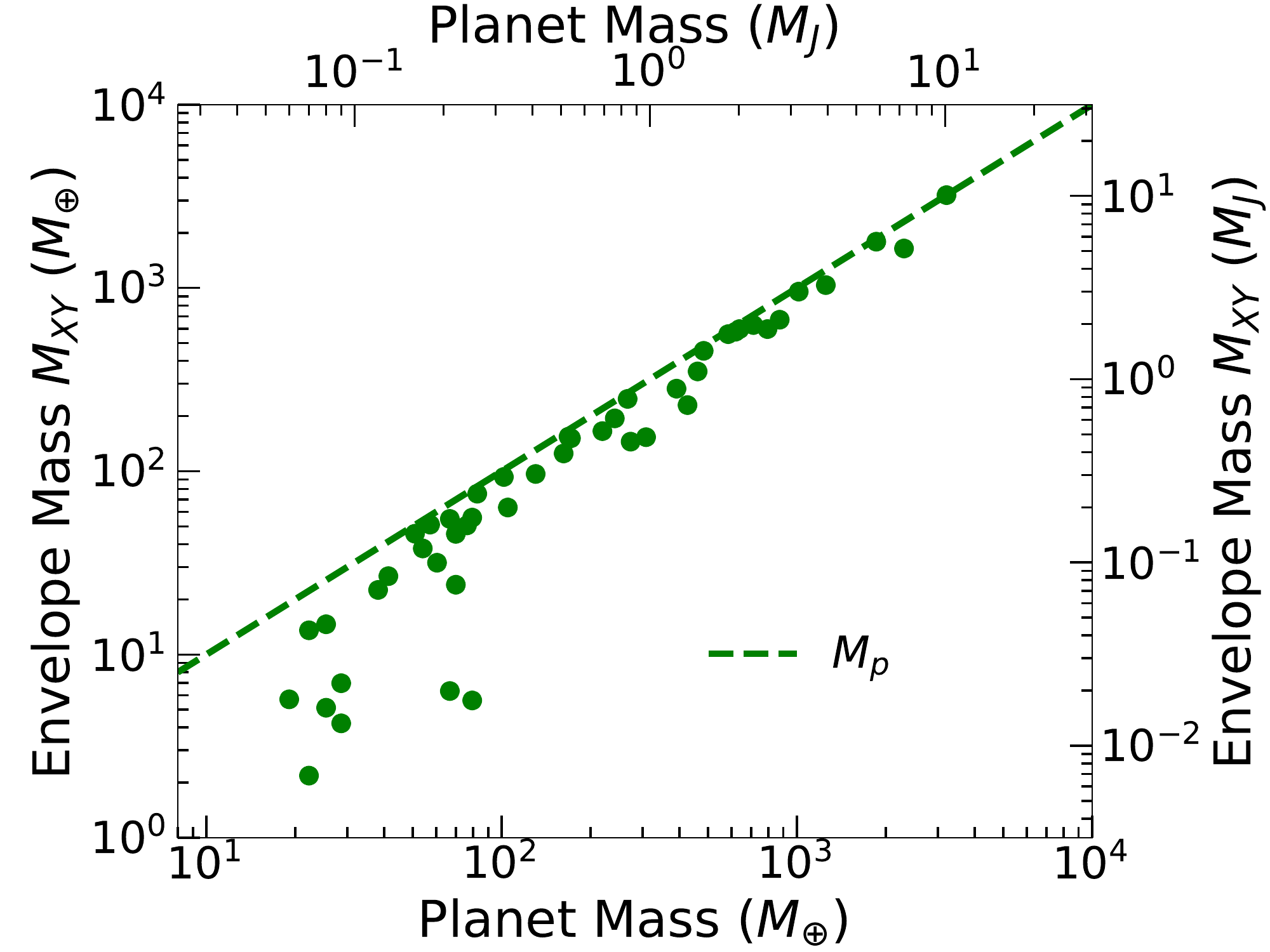}
\includegraphics[width=8cm]{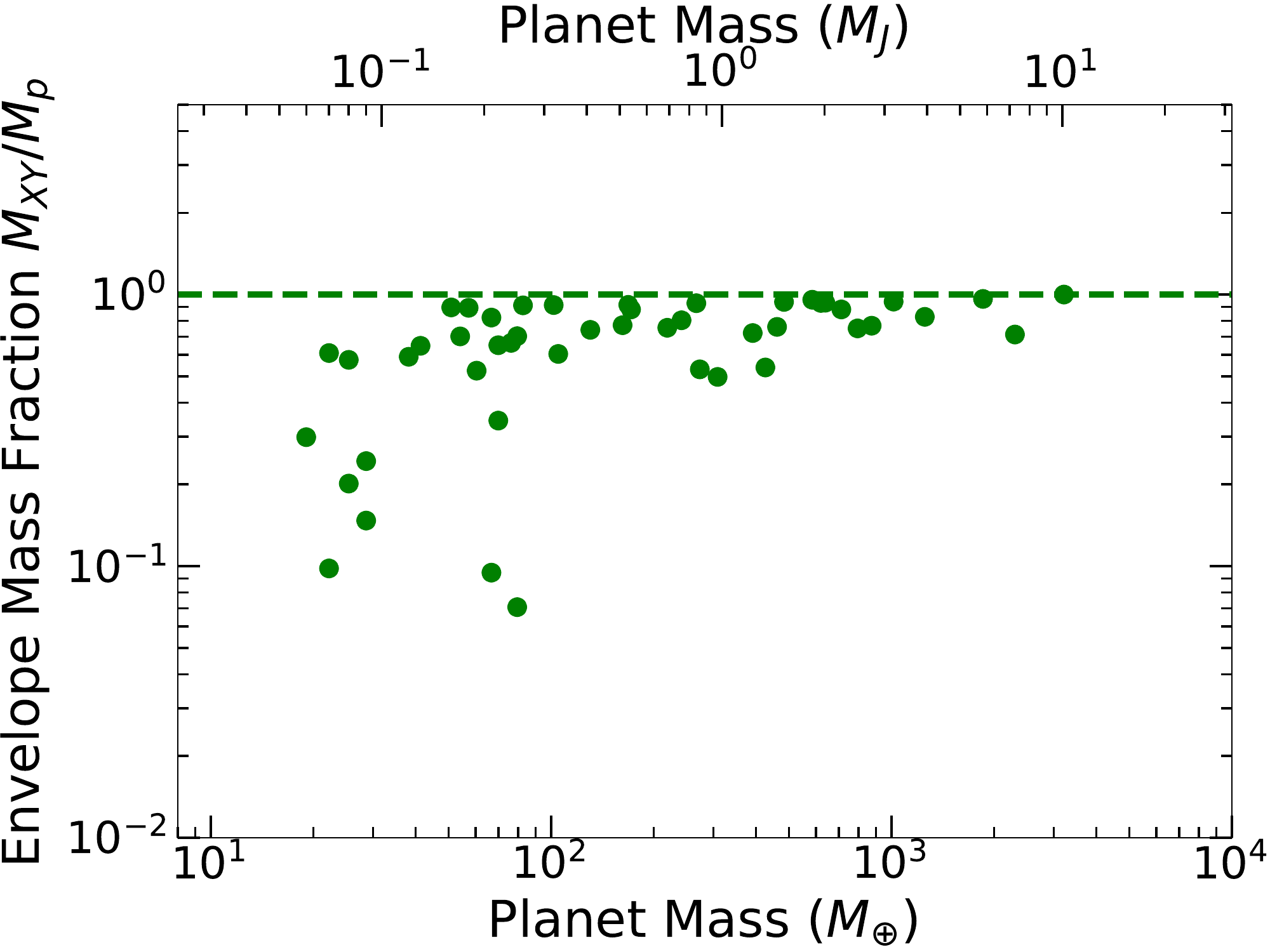}
\caption{The computed envelope mass of observed exoplanets as a function of planet mass.
Our reanalysis shows that most of the observed exoplanets experienced efficient gas accretion (see the green dots).
This trend is clearly seen on both plots of the envelope mass ($M_{XY}$ on the left panel) and of the mass fraction ($M_{XY}/M_{p}$ on the right panel).
On both panels, the green dashed line denotes the straight line of $M_{XY}=M_p$ for the reference.
It is interesting that some planets that have the mass of $\sim 20-100 M_{\oplus}$ have low values of $M_{XY}/M_{p}$,
indicating that they did not undergo runaway gas accretion.
The value of $\sim 20-100 M_{\oplus}$ is larger than the canonical value of the critical core mass that is about $10 M_{\oplus}$ in the literature.
Our simple calculations therefore suggest that efficient gas accretion tends to be postponed for some exoplanets.
}
\label{fig2}
\end{center}
\end{minipage}
\end{figure*}

Figure \ref{fig2} depicts the computed value of $M_{XY}(=M_p-M_Z)$ and the mass fraction ($M_{XY}/M_p$) as a function of $M_p$ on the left and right panels, respectively.
Our simple calculations show that the envelope mass becomes comparable to the total mass of planets 
when they are more massive than $\sim 100 M_{\oplus}$  (see the green dots on the left panel).
This suggests that efficient gas accretion occurred for all of the observed exoplanets that have masses larger than $\sim 100 M_{\oplus}$,
which is also confirmed by the mass fraction of $M_{XY}$ (see the right panel).
Importantly, we find that some of planets in the mass range of $20 M_{\oplus} \la M_{p} \la 100 M_{\oplus}$ did not experience efficient gas accretion.
Given that previous studies demonstrate that the critical core mass is about $5-10 M_{\oplus}$ (see Section \ref{sec:mod_1}), 
our computations indicate that some mechanisms would be needed to postpone the onset of efficient gas accretion 
for some exoplanets until their masses reach $\sim 20-100 M_{\oplus}$.
Note that the upper value of $M_p(\simeq 100 M_{\oplus})$ comes from only two points (see Figure \ref{fig2}).
This critical value may change when more and improved results of planet structure models would become available.

\subsection{The effect of solid accretion} \label{sec:data_3}

\begin{figure}
\begin{center}
\includegraphics[width=8cm]{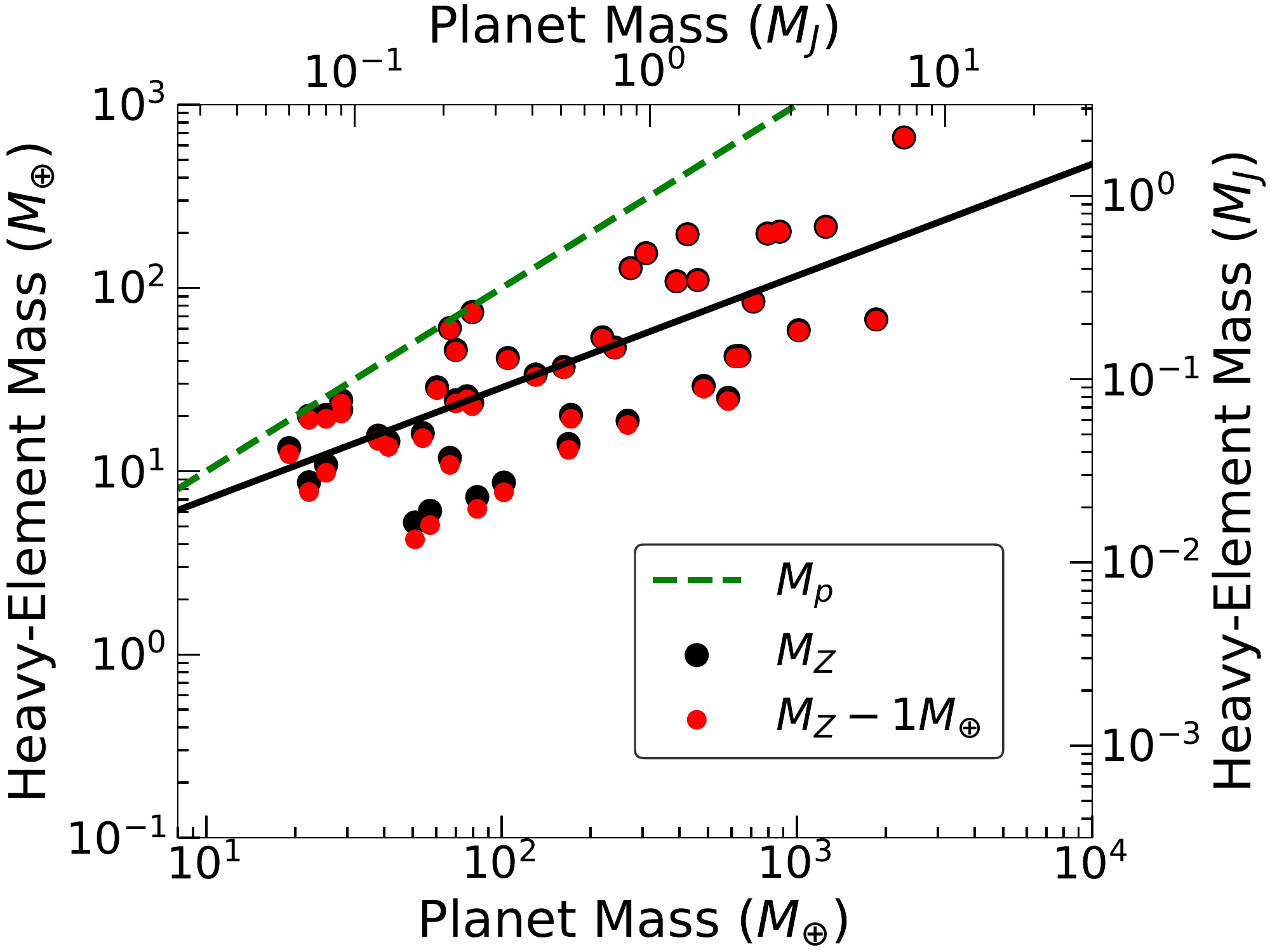}
\includegraphics[width=8cm]{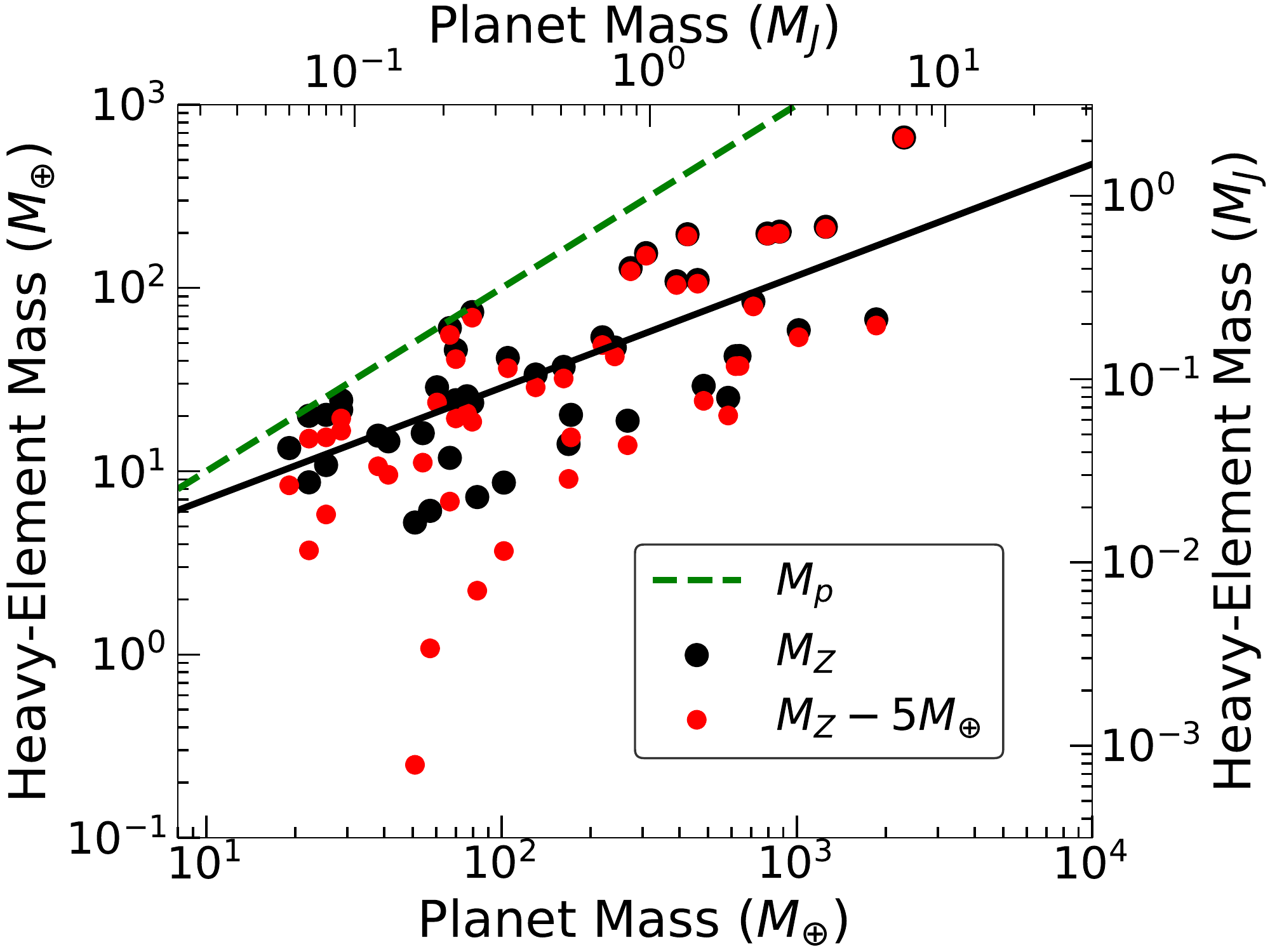}
\includegraphics[width=8cm]{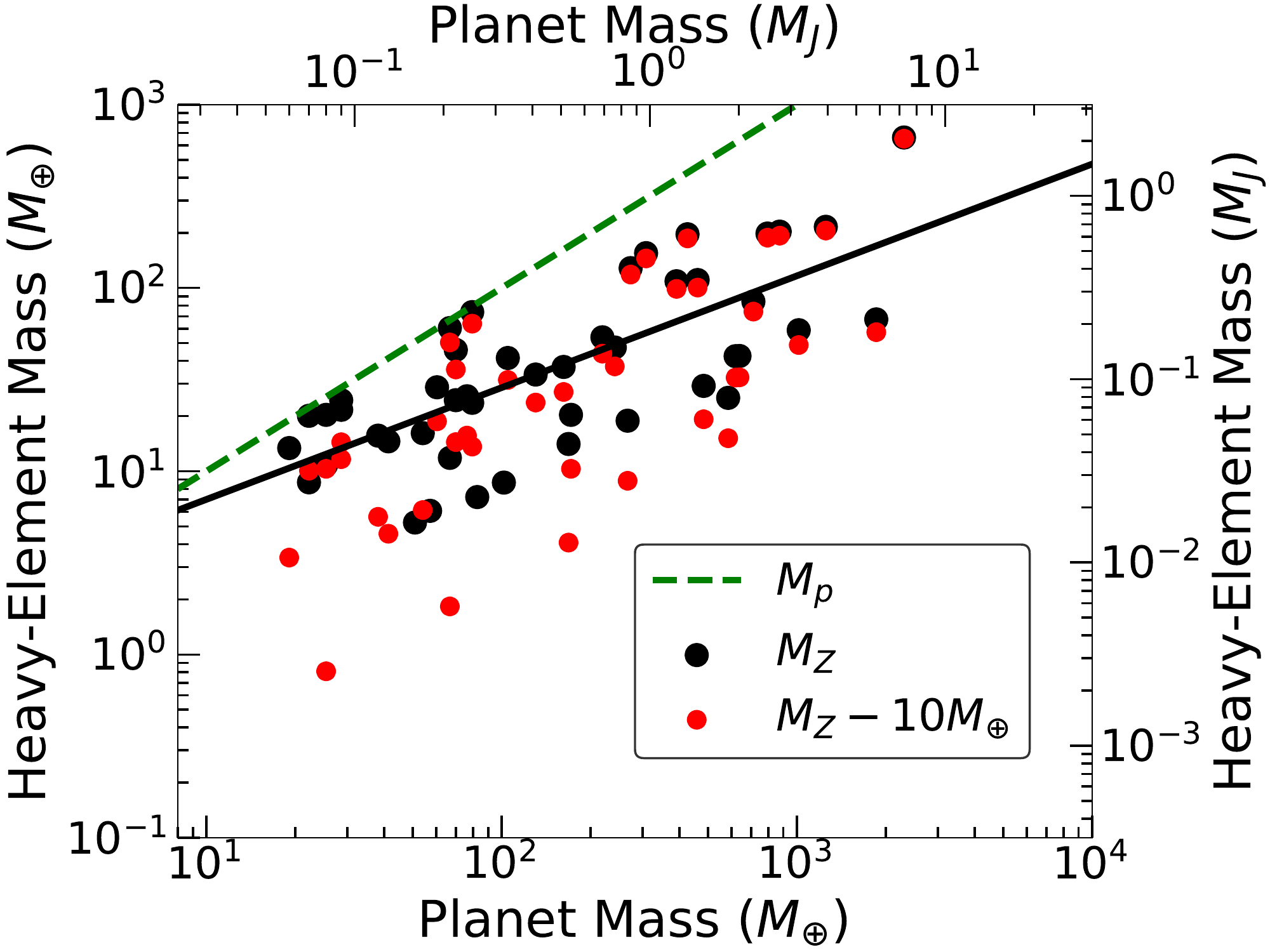}
\caption{
Heavy element mass as a function of $M_p$ for observed exoplanets. 
As in Figure \ref{fig1} (left), the computed values and the best fit derived by T16 are denoted by the black dots and the black solid line, respectively on each panel.
We also plot the straight line of $M_Z=M_p$  for reference (see the green dashed line).
From the top to the bottom, the assumed core mass ($1 M_{\oplus}$, $5 M_{\oplus}$, and $10M_{\oplus}$) is subtracted from $M_{Z}$, respectively.
This parameterized approach shows that the power-law index for the $M_Z-M_p$ relation tends to be smaller with increasing $M_p$.
This trend is well reproduced when observed exoplanets formed under the condition that gap formation is achieved in planetesimal disks around the planets 
and gas accretion onto the planets is controlled by disk evolution ($\tau_{g,acc}=\tau_{g,hydro}$, see Table \ref{table3}).
Our analysis therefore implies that the relationship discovered by T16 provides the useful information for the final stage of planet formation.
}
\label{fig3}
\end{center}
\end{figure}

\begin{figure}
\begin{center}
\includegraphics[width=8cm]{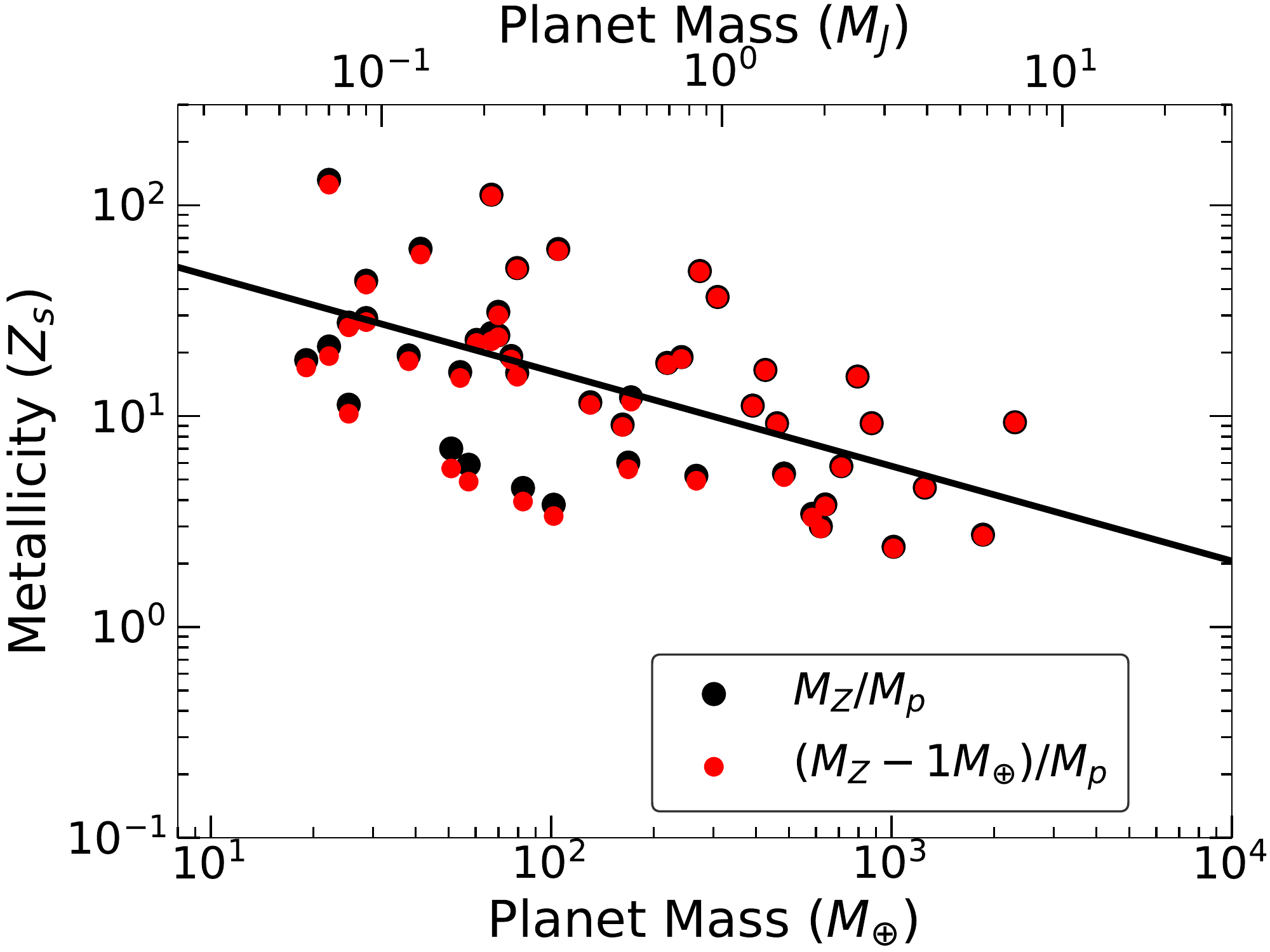}
\includegraphics[width=8cm]{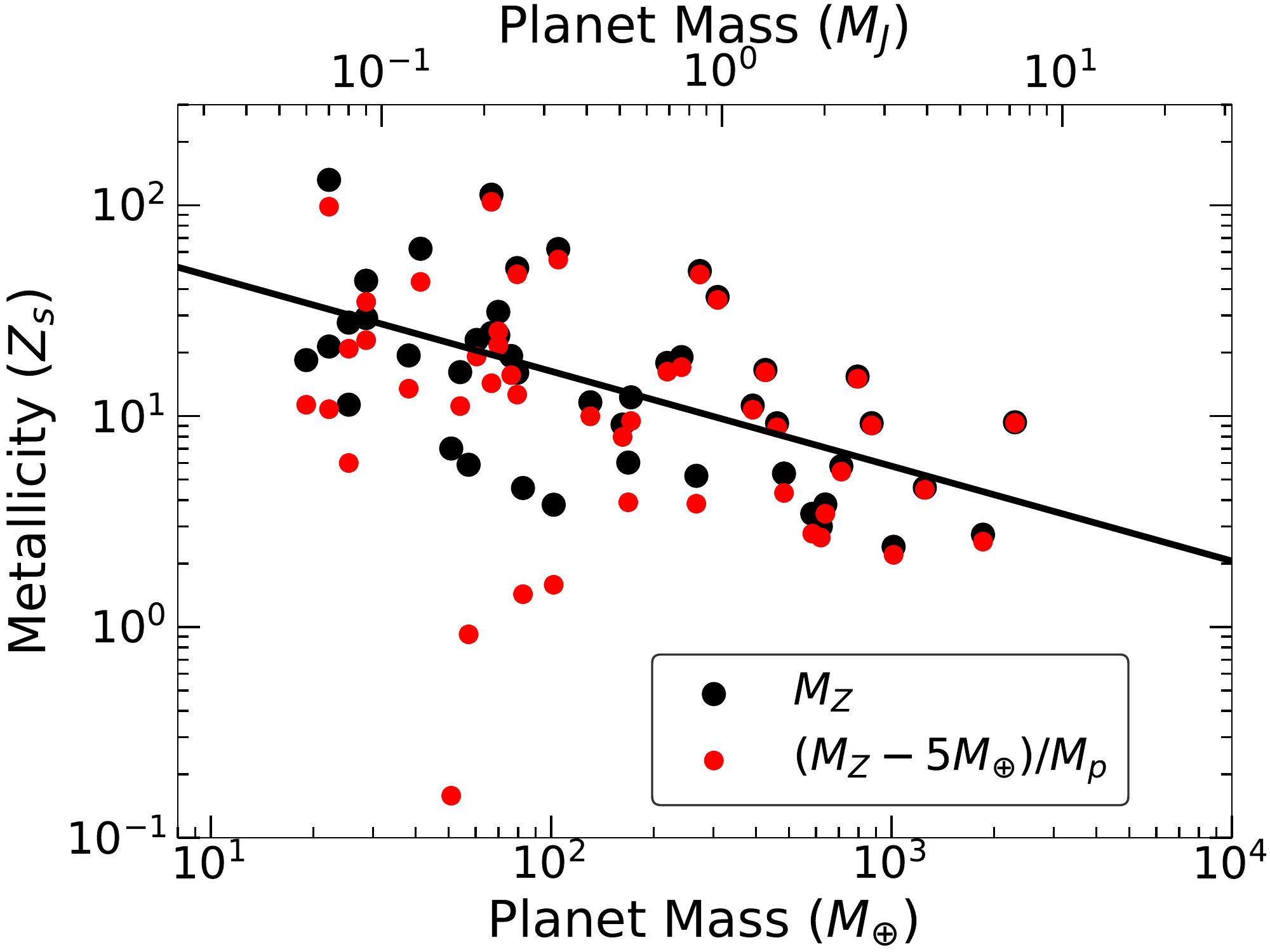}
\includegraphics[width=8cm]{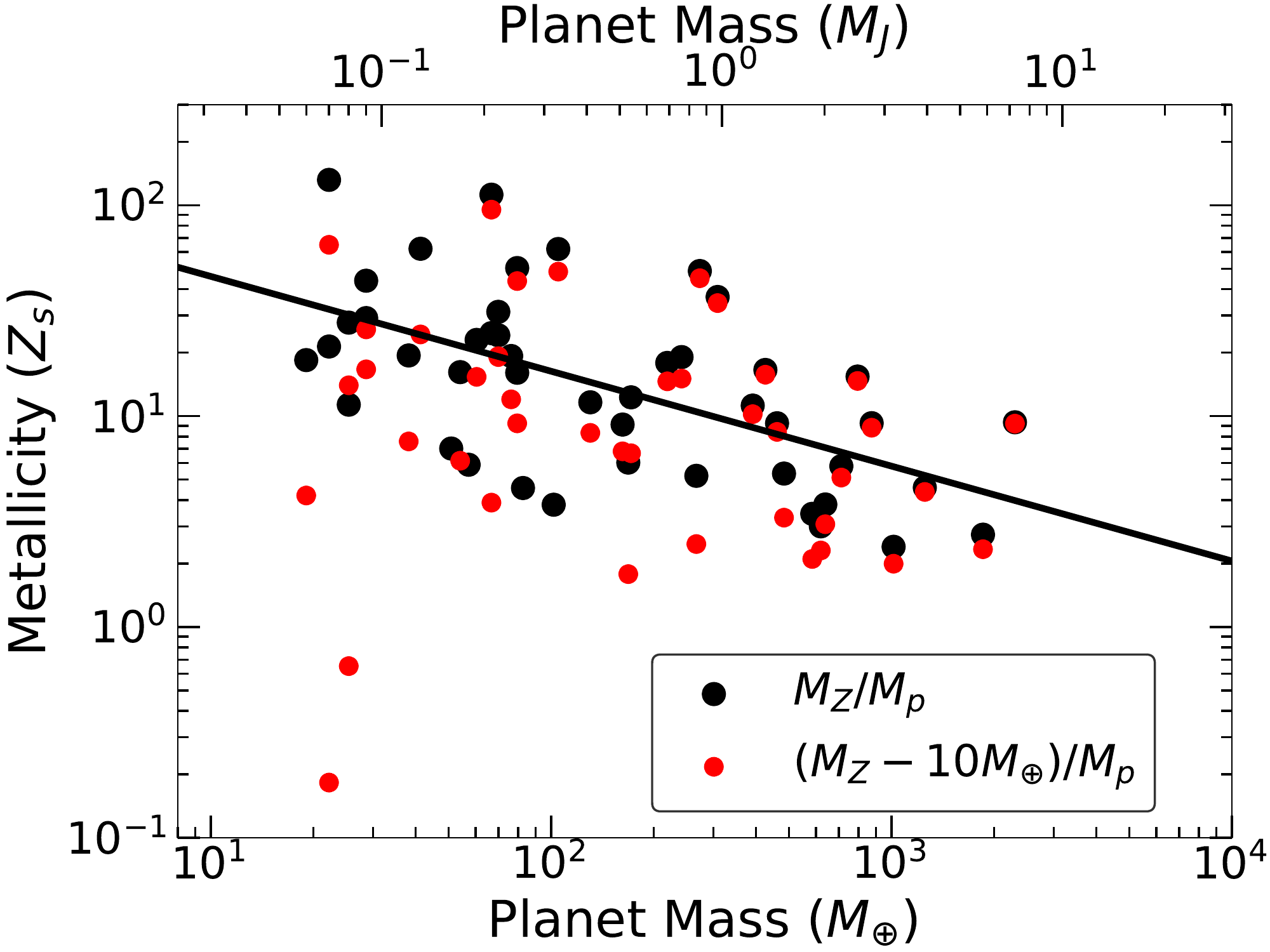}
\caption{
Metallicity as a function of $M_p$.
As in Figure \ref{fig1} (right), the computed values of $Z_p$ and the best fit of T16 are plotted as the black dots and the black solid line, respectively on each panel.
We again adopt the parameterized approach for the core mass,
in order to examine how subtraction of possible values ($1M_{\oplus}$, $5M_{\oplus}$, and $10 M_{\oplus}$) of the core mass 
affects the $Z_p-M_p$ relation from the top to the bottom panel, respectively (as done in Figure \ref{fig3}).
Under the assumption that the envelope metallicity of planets is purely determined by solid accretion in the post-core formation stage,
our results can be viewed that a correlation between envelope metallicity and planet mass should be characterized by a shallower slope.
Also, some transition in envelope metallicity should be present at the mass range of $10 M_{\oplus} \la M_{p} \la 100 M_{\oplus}$,
which may be related to the core mass.
}
\label{fig4}
\end{center}
\end{figure}

We here examine the effect of solid accretion on the $M_Z-M_p$ and $Z_p-M_p$ relations.
Given that solid accretion can divide into the core formation stage ($M_{core}$) and the post-core formation stage ($M_{pl}$ and $M_{pe}$, see equation (\ref{eq:mz_solid})),
we subtract $M_{core}$ from $M_{Z}$ and explore the resulting behavior of the heavy element mass ($= M_{Z}-M_{core}$).
Note that as discussed in Section \ref{sec:analysis_5}, the contribution of $M_{Z,gas}$ is negligible (also see Section \ref{sec:data_4}).

Figure \ref{fig3} shows the results of our analysis.
Since it is unknown what is the initial core mass for these planets,\footnote{T16 treated the core mass as a free parameter with the upper limit of $10 M_{\oplus}$, 
and their best fit values are not provided in their paper.}
we adopt a parameterized approach. 
In this approach, three plausible values ($1 M_{\oplus}$, $5 M_{\oplus}$, and $10 M_{\oplus}$) of the core mass are subtracted. 
We find that as the subtracted core mass increases (from the top to the bottom panel of Figure \ref{fig3}), 
the slope of the heavy element mass becomes steeper, especially at the less massive ($M_p \la 10^3 M_{\oplus}$) region (see the red dots).
This is simply because when planets are not so massive, the total heavy element mass is also relatively small.
If a certain value of the core mass is removed from $M_{Z}$, 
then the reduction in $M_{Z}$ becomes more enhanced for lower mass planets than massive ones.
Thus, our analysis indicates that the slope tends to be steeper ($> 3/5$) for planets with the mass of $\la 10^3 M_{\oplus}$ and to be shallower ($\simeq 3/5$) for more massive planets
when the core mass is subtracted from the total heavy element mass ($M_{Z}$).

We now turn our attention to the $Z_p-M_p$ relation.
For this case, we utilize the results of our analysis to develop an interpretation that is different from the above one.
More specifically, we assume that the metallicity computed from $M_Z-M_{core}$ represents the envelope metallicity.
This assumption would be valid if planetary cores do not dissolve into their envelopes and 
if solids accreted in the post-core formation stage fully dissolve into the envelopes due to thermal ablation.

Figure \ref{fig4} shows the results. 
We have adopted the same parameterized approach as above.
From top to bottom,
the assumed core mass that is removed from $M_Z$ is altered from $1 M_{\oplus}$, $5 M_{\oplus}$, and $10 M_{\oplus}$, respectively.
Our analysis shows that subtraction of the core mass from $M_Z$ tends to wash out the $Z_p-M_p$ relation,
especially for planets that have masses of $<20-100 M_{\oplus}$ (see the red dots).
This occurs simply because the value of planetary metallicity ($Z_p$) is more affected for lower-mass planets, as discussed above.
If the above assumption would be reasonable for observed exoplanets and envelope metallicity is determined only by the subsequent solid accretion,
then our results can be interpreted that a correlation between envelope metallicity and planet mass should have a shallower slope than that of the $Z_p-M_p$ relation.
Also, there should be a transition in envelope metallicity as the planet mass increases.
This transition would be related to the core mass.
In Section \ref{sec:disc_3}, we will discuss more about how these interpretations are related to the current observations of exoplanets' atmospheres.

\subsection{The effect of gas accretion} \label{sec:data_4}

\begin{figure*}
\begin{minipage}{17cm}
\begin{center}
\includegraphics[width=8cm]{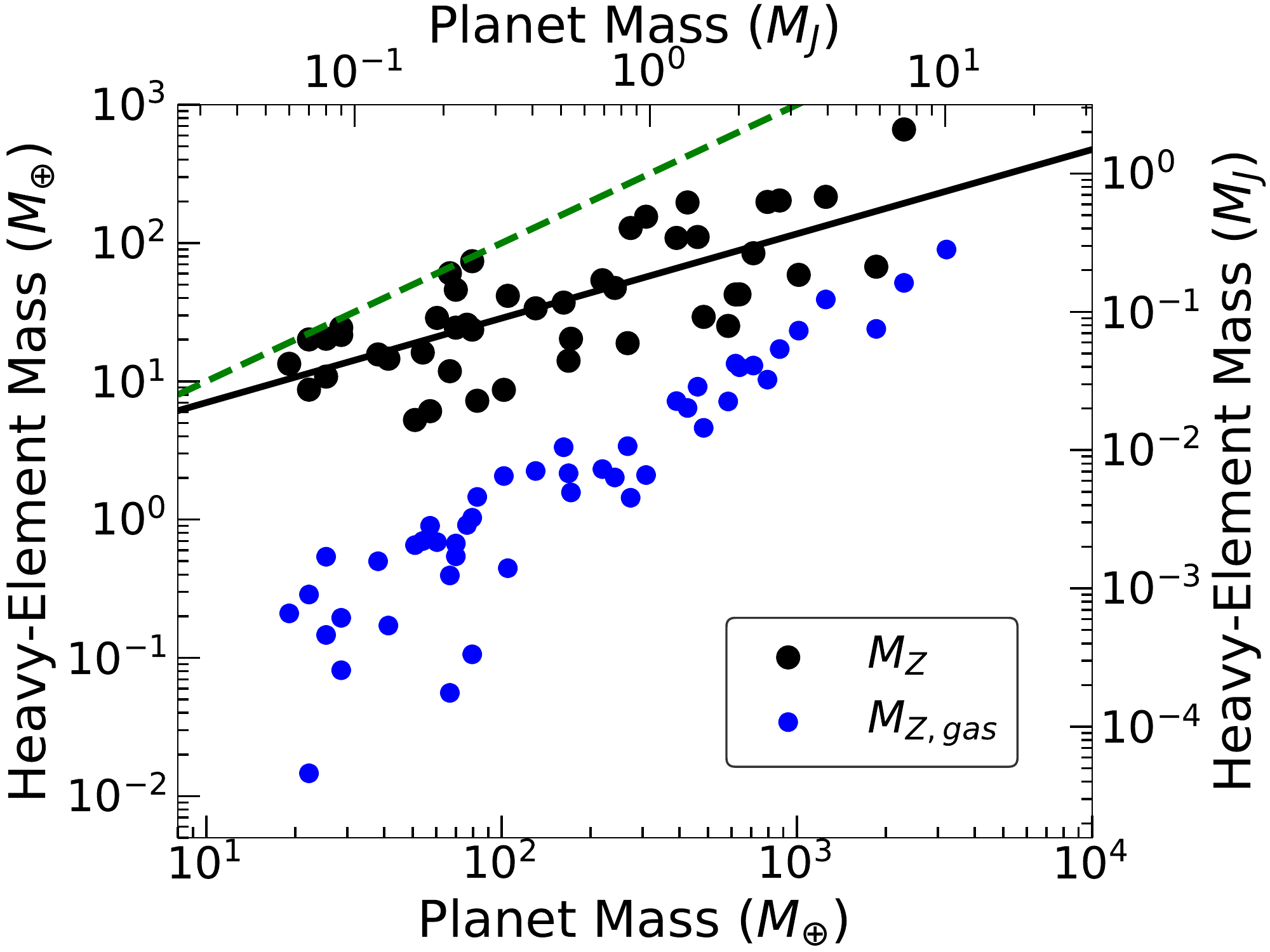}
\includegraphics[width=8cm]{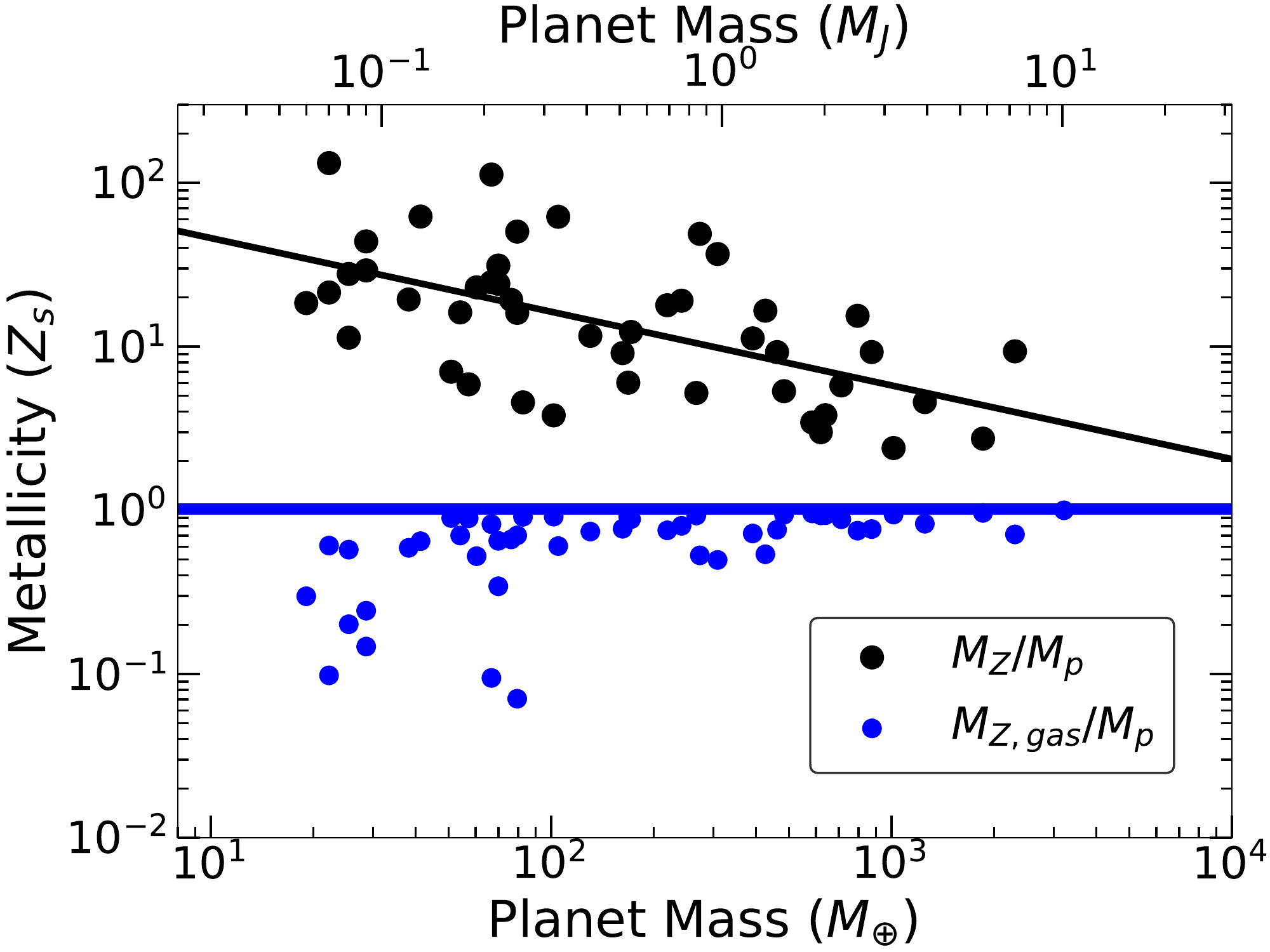}
\caption{
Gas accretion and its contribution to $M_Z$.
As in Figure \ref{fig1}, the black dots and the black solid line represent the estimated values of T16 and its best fit, respectively.
For comparison purpose, the straight line of $M_Z = M_p$ is denoted by the green dashed line on the left panel,
and the straight line of $Z_p = Z_s$ is by the blue solid line on the right panel.
Our analysis shows that the computed value of $M_{Z, gas} (=Z_s M_{XY})$ is much smaller than that of $M_Z$ (see the blue dots).
This indicates that heavy elements that are accreted following gas accretion are not crucial for the value of $M_Z$
until the planet mass exceeds $\ga 10^3 M_{\oplus}$.
Also, we confirm that the disk gas accreted onto planets contains the dust abundance that is similar to the stellar metallicity (see the blue line).
}
\label{fig5}
\end{center}
\end{minipage}
\end{figure*}

We here consider the effect of gas accretion ($M_{Z,gas}$) on the total heavy element mass ($M_Z$) and the planet metallicity ($Z_p$).

As already shown in Section \ref{sec:analysis_5}, the contribution of $M_{Z,gas}$ is readily computed for given values of $M_Z$, $M_p$, and $Z_s$ (see equation (\ref{eq:mz_gas_mz_solid})).
Figure \ref{fig5} shows the resulting values (see the blue dots).
Our analysis confirms that dust accretion accompanying with gas accretion is not crucial for understanding the total heavy element mass of observed planets (see the left panel).
We also find that the contribution of dust accretion is an order of unity for massive ($\ga 100 M_{\oplus}$) planets (see the right panel).
This can be viewed as a verification of the assumption that the dust abundance in the accreted gas is about $Z_s$.

\subsection{Comparison with our theoretical analysis} \label{sec:data_5}

\begin{table*}
\begin{minipage}{17cm}
\begin{center}
\caption{Summary of power-law indices of $M_{Z}(\propto M_p^{\Gamma})$ and $Z_{p}(\propto M_p^{\beta})$}
\label{table3}
\begin{tabular}{l|c|c|c|c|c|c} 
\hline 
Power-law index     & Planetesimal Accretion                           & Planetesimal Accretion                                       & Pebble Accretion                                   & T16                        & M14$^a$   & KB14$^b$     \\ 
                               & No Gap                                                   & Gap                                                                     &                                                                &                              &             &              \\  \hline 
$\Gamma$             & $- 2/5 $ ($\tau_{g,acc}=\tau_{g,KH}$)    & $ \simeq 2$ ($\tau_{g,acc}=\tau_{g,KH}$)           & $- 10/3 $ ($\tau_{g,acc}=\tau_{g,KH}$)  & $0.61 \pm 0.08 $  &             &              \\ 
                               & $1/3$ ($\tau_{g,acc}=\tau_{g,hydro}$)   &$ \simeq 3/5$ ($\tau_{g,acc}=\tau_{g,hydro}$)     & $1/3$ ($\tau_{g,acc}=\tau_{g,hydro}$)   & $\simeq 3/5$         &             &              \\
$\beta$                   & $- 7/5$ ($\tau_{g,acc}=\tau_{g,KH}$)     & $ \simeq 1$ ($\tau_{g,acc}=\tau_{g,KH}$)           & $- 13/3$ ($\tau_{g,acc}=\tau_{g,KH}$)   & $-0.45 \pm 0.09$  & $-0.68$ & $-1.1$   \\
                               & $- 2/3$ ($\tau_{g,acc}=\tau_{g,hydro}$) & $ \simeq - 2/5$ ($\tau_{g,acc}=\tau_{g,hydro}$)  & $- 2/3$ ($\tau_{g,acc}=\tau_{g,hydro}$) & $\simeq -2/5$       & $\simeq -2/3$ &   \\
\hline 
\end{tabular}
\end{center}
$^a$ see equation (\ref{eq:m14}).

$^b$ see equation (\ref{eq:kb14}).
\end{minipage}
\end{table*}

We now compare our theoretical results (see Section \ref{sec:analysis}) with those of T16 (see Figure \ref{fig1}).
To proceed, we summarize our results in Table \ref{table3}.

We find that if the core mass of observed exoplanets is relatively small ($\la 1 M_{\oplus}$),
the best fit is achieved for
the case where planetesimal accretion is slowed down due to gap formation and gas accretion is also limited by disk evolution (see Figure \ref{fig3} and Table \ref{table3}).
This implies that the $M_{Z}-M_p$ relation would be determined predominantly by the final stage of planet formation.
Even if the core mass of these planets would be relatively large ($\simeq 5-10 M_{\oplus}$),
the trend for observed exoplanets can be reproduced well only in the case of gap formation in planetesimal disks (see Section \ref{sec:data_3}):
as the value of $M_p$ increases, 
the slope for the correlation between the heavy element mass and $M_p$ becomes shallower with increasing the planet mass.
Mathematically, the power-law index changes from $2$ to $3/5$ for this case (see Table \ref{table3}).

As a conclusion, our analysis suggests that 
the trend found by T16 would be understood well if it traces the final stage of planet formation:
Planets are already massive enough to generate a gap in their surrounding planetesimal disks 
and the gas accretion rate onto the planets is considerably reduced and mainly regulated by disk evolution.

\section{Discussion} \label{sec:disc}

We first list up the limitations of our analysis.
We then discuss other physical processes that are not considered in the above analyses, 
and examine their effects on our conclusions.
Also, we summarize previous studies which are directly related to this work,
and compare them with our finding.
We provide a comprehensive picture of planet formation that is derived from our analysis, 
and finally discuss some implications for the current and future observations of exoplanets and their atmospheres.

\subsection{Limitation of our analysis} \label{sec:disc_1}

We here discuss the limitations of our analysis.

The first limitation is that the trend discovered by T16 is based on only 47 exoplanets (see equations (\ref{eq:m_z_obs}) and (\ref{eq:z_p_obs})).
It is well known that while hot Jupiters are statistically rare, actually observed planets are not rare 
since they are readily observed by both radial velocity and transit methods \citep[e.g.,][]{wf15,dj18}.
This limitation indeed originates from an incomplete understanding of inflation mechanisms of hot Jupiters (T16).
Once the dominant mechanism is identified, a similar analysis will be carried out to such hot Jupiters.
Furthermore, the current and future observations attempt to improve measurements of both the mass and radius of detected exoplanets.
Such better data will make it possible to apply a similar analysis not only to hot/warm Jupiters but also to smaller sized planets.
Thus, it is currently not obvious that the $M_Z-M_p$ and the $Z_p-M_p$ relations derived by T16 are universal for various types of planets,
which remains to be explored in the future work.

The second limitation is that our analysis heavily relies on the computed value of $M_Z$.
As discussed in Section \ref{sec:data_1}, both better observational data and modeling are needed to constrain the value of $M_Z$ tightly.
T16 pointed out that the present observational data are still not good enough (see their section 4.1).
As a result, the error bars of $M_Z$ are currently determined mainly by uncertainties in mass and radius measurements of observed exoplanets.
Even if the observational data become better, uncertainties in model parameters cannot be fully removed.

The third limitation is involved with our approach.
In this approach, we focus only on the power-law indices of the $M_Z-M_p$ and the $Z_p-M_p$ relations,
in order to elucidate the underlying physics.
This simplification needs to be examined carefully by detailed numerical simulations.
In particular, recent studies show that the gas accretion process behaves differently with different assumptions and numerical setups \citep[e.g.,][]{mim10,db13,vab16,ll17}.
We however emphasize that our adopted formula fits well the results of numerical simulations that are performed by different groups such as \citet{tw02,dkh03,mim10}.
As clearly shown in figure 1 of  \citet{tt16}, 
the formula works well for planets with the mass range of $10 M_{\oplus} \la M_p \la 30 M_{\oplus}$ with a specific disk model
that has the gas surface density of $140$ g cm$^{-2}$, the aspect ration of 0.05, and the turbulent parameter $\alpha$ of $4 \times 10^{-3}$ \citep{ss73}
at the planet position of $r=5.2$ au. 
This implies that once gas accretion is regulated by disk evolution (see equation (\ref{eq:tau_hydro})), 
the formula would become reasonable until a (clear) gap is curved in gas disks.
Note that \citet{lhdb09} investigate gas accretion onto planetary cores, taking into account disk-planet interaction.
While they derive a different form of the gas accretion recipe (see their equation (2)),
they adopt simulations of \citet{dkh03}.
Thus, our formula should be broadly consistent with theirs.
A severer limitation of our approach is that we cannot compute the absolute value of $M_Z$ directly.
The value would be determined by the combination of model and disk parameters.
We will leave such a detailed study for the future work.

\subsection{Other physical processes} \label{sec:disc_2}

In this section, we consider the effect of other physical processes that are not included in our analyses.
These include orbital evolution due to planetary migration, gas gap formation by the migration, and the effect of nearby forming planets.

First, we point out that our analyses do not take into account the orbital evolution of planets by planetary migration \citep[e.g.,][]{kn12}.
It is expected that planetary migration allows protoplanets to replenish planetesimals in their feeding zones.
This is because the protoplanets can sweep up a new region of their planetesimal disks.
In fact, \citet{ambw05} show that migrating protoplanets can have more chance to accrete a larger number of planetesimals in the disks,
which speeds up core formation.
It is however important to emphasize that more detailed simulations with a direct $N-$body integrator suggest that 
the planetesimal accretion rate and gap formation in planetesimal disks depend on the migration speed, which is a function of planet mass \citep{ti99}.
A more self-consistent simulation is needed to investigate how gaps form around growing, migrating planets in their planetesimal disks,
and how semi-analytical formulae can be affected due to planetary migration (see equations (\ref{eq:si08_24}) and (\ref{eq:si08_25})).

Second, we discuss gap formation in gas disks that is the inevitable outcome of migration, especially for massive planets \citep[e.g.,][]{npmk00,cmm06,hi13,dk15}.
As described in Sections \ref{sec:analysis_1}, the $M_Z-M_p$ relation is determined not only by solid accretion, but also gas accretion onto planets (see equation (\ref{eq:dmz_dmp})).
In the above analyses, the effect of gas gaps has not been considered explicitly.
This is because our analyses heavily rely on the results of \citet{si08},
and their results are obtained under the assumption of no gap formation in gas disks for simplicity.
One might consider that the presence of gas gaps would affect our conclusion very much 
since the gas surface density can now become a function of planet mass (see $\sigma_{p,acc}$ in equation (\ref{eq:tau_hydro})).
In order to address this point, we develop a similar analysis in Appendix \ref{app1}. 
Here we briefly summarize the results.
We find that the trend found by T16 can be reproduced 
only when gaps are present in gas disks but no gap in planetesimal disks (see Table \ref{tableA1}).
We argue that this situation is very unlikely to be achieved.
This is because gap formation takes place more readily in planetesimal disks than gas disks due to the lack of the pressure term.
Furthermore, even if planets accrete gas and solids from gapped gas disks,
the total amounts of accreted gas and solids would not be significant, compared with those accreted from gas disks without any gap \citep[e.g,][]{ti07,tt16}.
Accordingly, it would be reasonable to consider that the trend of $M_{Z}$ is determined predominantly before gap formation takes place in gas disks 
and such a trend does not change very much after gas gap formation.
Thus, our conclusion would be maintained even if gap formation in gas disks is properly taken into account,
while a more self-consistent simulation is needed to fully justify this consideration.

Third, we have so far assumed implicitly that planet formation proceeds in an isolated region,
that is, we consider formation of single planets.
We must admit that this is a highly idealized situation.
In reality, multiple planets form in single disks at the same time,
and the gravitational interaction arising from nearby growing planets would affect the dynamics of planetesimals there.
This can change the spatial distribution of planetesimals and hence the condition of gap formation in planetesimal disks.
It is interesting to investigate how the $M_{Z}-M_p$ relation can be altered 
when formation of multiple planets is considered appropriately.

Thus, while some improvements would be required in our analyses for developing a more complete picture of planet formation,
our present results are still useful for understanding a number of the currently known observational trends.

\subsection{Comparison with previous studies} \label{sec:disc_3}

\begin{figure}
\begin{center}
\includegraphics[width=8cm]{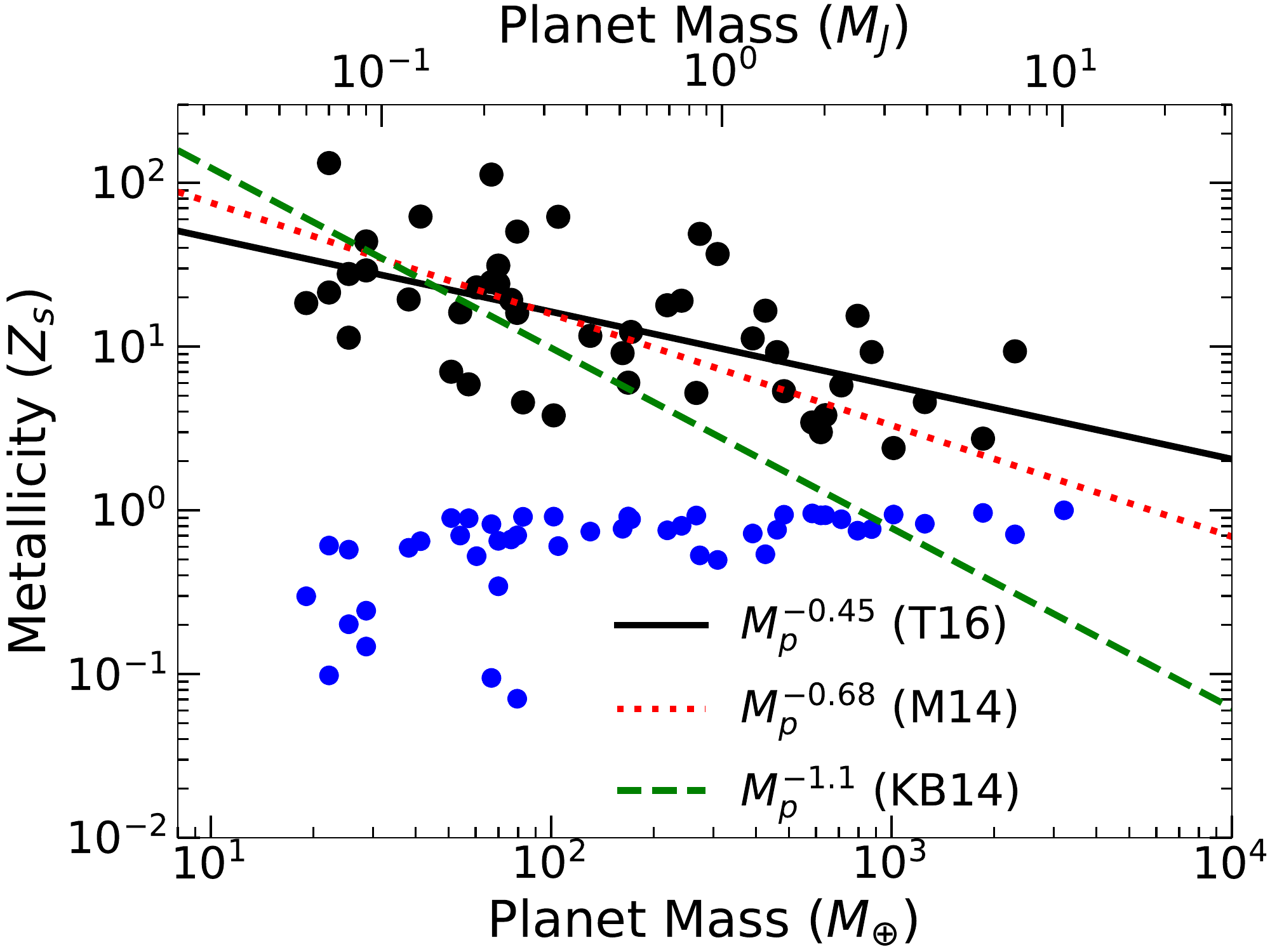}
\caption{Comparison with previous studies.
As in Figure \ref{fig5}, the black dots and the black solid line represent the results of T16 and its best fit, respectively.
For comparison purpose, the results of \citet{mka14} and of \citet{kbd14} are denoted by the red dotted and the green dashed lines, respectively.
Note that the former computes the total heavy element mass while the latter is for atmospheric metallicities.
It is interesting that the slope of M14 is well reproduced by our simple analysis when the same assumption is adopted,
that is, solid accretion proceeds from planetesimal disks without any gap and gas accretion is limited by disk evolution (see Table \ref{table3}).
The slope of KB14 is the most steepest.
This may suggest that dust grain growth and settling is more efficient for more massive planets.
In other words, a difference in slopes between the total heavy element mass (the black solid line) and the atmospheric metallicity (the green dashed line) 
may be used as a tracer of metallicity evolution of exoplanets' atmospheres.}
\label{fig6}
\end{center}
\end{figure}

In this section, we touch on recent studies that are relevant to this work and compare their findings with ours.

One of the most advanced models that compute the total heavy element mass in planets are \citet{mka14,mvm16}.
In this model, the standard core accretion picture is adopted to trace mass growth of planets.
By coupling with planetary migration, 
they also make use of an enhanced planetesimal accretion rate, following the approach of \citet{ambw05}.
While they do not treat dust physics in planetary envelopes self-consistently,
they mimic this effect by artificially reducing the grain opacity there \citep{mka14}.
Covering a large parameter space and performing population synthesis calculations, 
they find that the $Z_{p}-M_p$ relation is given as \citep[see Table 7 in][]{mvm16}
\begin{equation}
\label{eq:m14}
\left( \frac{Z_p}{Z_s} \right)_{M14} = 7.2 \left( \frac{M_p}{M_J} \right)^{-0.68},
\end{equation}
where $M_J$ is the Jupiter mass.
Note that this relationship is derived from the total heavy element mass \citep[$M_Z$, see][]{mka14}.
It is interesting that this slope is steeper than the results of T16 (see Figure \ref{fig6}, also see table \ref{table3}).
As discussed in Section \ref{sec:analysis_2}, 
the slope is regulated by both planetesimal dynamics and gas accretion onto planets.
In their model, the disk-limited gas accretion ($\tau_{g,acc}=\tau_{g,hydro}$) is taken into account, 
but the effect of gap formation in planetesimal disks is not.
As a result, their simulations lead to a steeper slope.
In fact, our analysis predicts the value of their slope, which is about $-2/3$ (see the case of no planetesimal gap in Table \ref{table3}).
Thus, \citet{mka14,mvm16} undertook a pioneering work and 
indicate the importance of planetesimal accretion for understanding the $Z_{p}-M_p$ relation.
And our follow-up work reproduces the results of T16 better and 
derive a clearer view of how the $Z_p-M_p$ relation can be used for obtaining better understanding of planet formation.

While we focus mainly on the total heavy element mass ($M_Z$) in this paper, 
it would be interesting to consider atmospheric metallicity as done in Section \ref{sec:data_3} (see Figure \ref{fig4}).
To proceed, we here discuss a correlation between envelope/atmospheric metallicity and planet mass.
As an example, we adopt the result of \citet{kbd14},
which is given as \citep[see Table \ref{table3}, also see Table 7 of][]{mvm16}
\begin{equation}
\label{eq:kb14}
\left( \frac{Z_p^{atm}}{Z_s} \right)_{KB14} = 2.75 \left( \frac{M_p}{M_J} \right)^{-1.1}.
\end{equation}
In their work, the metallicity of a hot Jupiter's atmosphere is estimated based on the precise determination of the water abundance in the atmosphere.
Combining the data points of four giant planets in the solar system, 
they obtain the above trend (see the green dashed line in Figure \ref{fig6}).
It is obvious that their slope is much steeper than that of T16.
Since such a steep slope cannot be explained by removing the initial core mass (see Figure \ref{fig4}),
we propose that the results of \citet{kbd14} are very likely to trace the metallicity evolution in exoplanet atmospheres:
dust grain growth and settling take place in planetary atmospheres, namely, in the top, thin layer of planetary envelopes,
and their effects are more pronounced for massive planets.
If this would be the case, comparison between the total heavy element mass ($M_Z$) and atmospheric metallicity can be used 
as an indicator of how atmospheric metallicity of planets evolves with time. 
Given that most of heavy elements should be present in planetary envelopes for massive planets, not in the core (see Figure \ref{fig3}),
they would be kept in the inner region of these envelopes.
It is interesting that numerical simulations already show that these processes operate efficiently in planetary envelopes
even during the process of forming \citep[e.g.,][]{mp08,mbp10}.
Note that the primordial envelope of planets should be more tenuous than the present one due to larger sizes,
which principally leads to inefficient dust growth and settling there.

Finally, we discuss pebble accretion.
As already pointed out in Section \ref{sec:analysis_3}, 
recent studies focus mainly on core formation \citep[e.g.,][]{blj15,jl17},
and application of their results to the final stage of planet formation may not be reasonable.
In fact, we find that the resulting power-law profile of the $M_Z-M_p$ relation is not consistent with the result of T16 (see Table \ref{table3}.)
It is nonetheless important to point out that there is significant potential that pebble accretion may play a role in understanding the $M_Z-M_p$ and $Z_p-M_p$ relations.
For example, it can be anticipated that
a large amount of pebbles would accumulate at the outer edge of the gas gaps after core formation is nearly terminated due to gas gap formation.
If this would be the case, such accumulation of pebbles would lead to planetesimal formation there.
Then, it would be possible to trigger the subsequent planetesimal accretion onto planets, 
which can eventually achieve enrichment of heavy elements in the planets.
In fact, high abundance of heavy elements in observed exoplanets requires a large amount of supplies that can potentially be delivered to
the feeding zone of planets via radial drift of pebbles.
Thus, while new numerical simulations of pebble accretion are desired,
pebble accretion might not play a direct role at the final stage of planet formation.

\subsection{A comprehensive picture} \label{sec:disc_4}

\begin{table*}
\begin{minipage}{17cm}
\begin{center}
\caption{Classification of observed exoplanets and the key physical processes of forming these planets}
\label{table4}
\begin{tabular}{l|c|c|c} 
\hline 
Name                                                & Mass range                                                                      & Color in Figure \ref{fig7}   & Key process$^a$                                       \\  \hline \hline
Rocky planets                                  & $ M_p \la  4 M_{\oplus}$                                                   & Blue                                 & Significant solid accretion with                     \\  
(or (super)Earth-type)                      &                                                                                           &                                         & an almost negligible amount of gas            \\ \hline
Gas-poor sub-giants                        &  $4M_{\oplus} \la M_p \la  100 M_{\oplus}$                      & Green                              & Planetesimal accretion with a gap    \\
(or Neptune-type)                            &                                                                                            &                                         & \& slowed-down gas accretion                     \\ \hline
Gas-rich giants                                & $100M_{\oplus} \la M_p \la  3 \times 10^4 M_{\oplus}$     & Grey                                & Planetesimal accretion with a gap                \\
(or Jovian-type)                               & ($0.4 M_{J} \la M_p \la  10^2 M_{J}$)                                &                                         & \& slowed-down gas accretion                      \\ \hline
Stars                                                & $10^2 M_{J} \la M_p $                                                       & Yellow                              & Collapse of self-gravitating gas                     \\
\hline 
\end{tabular}

$^a$ Detailed numerical simulations and further modeling for the observations of exoplanets are obviously needed to confirm our prediction.
\end{center}
\end{minipage}
\end{table*}

\begin{figure*}
\begin{minipage}{17cm}
\begin{center}
\includegraphics[width=8cm]{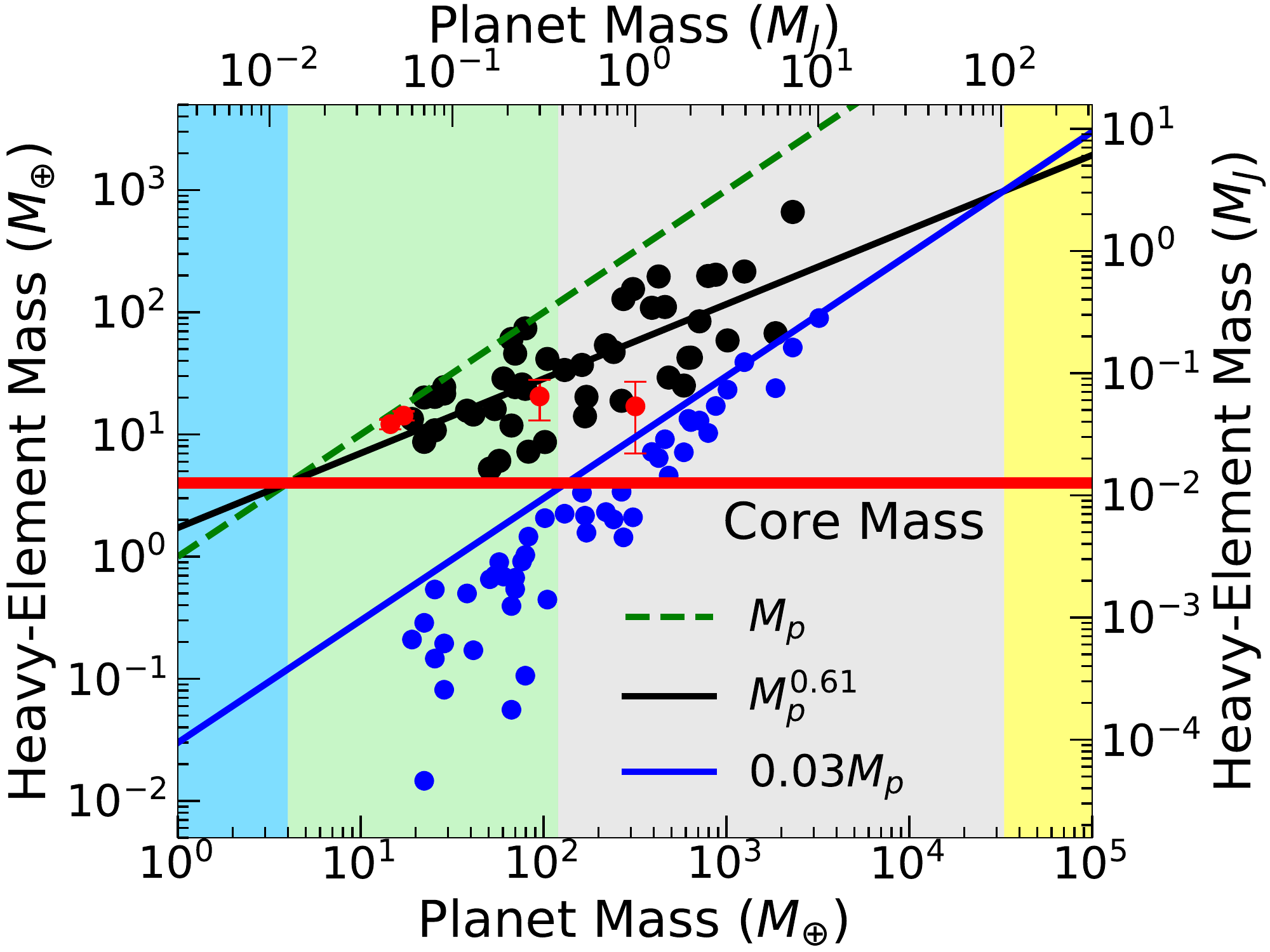}
\includegraphics[width=8cm]{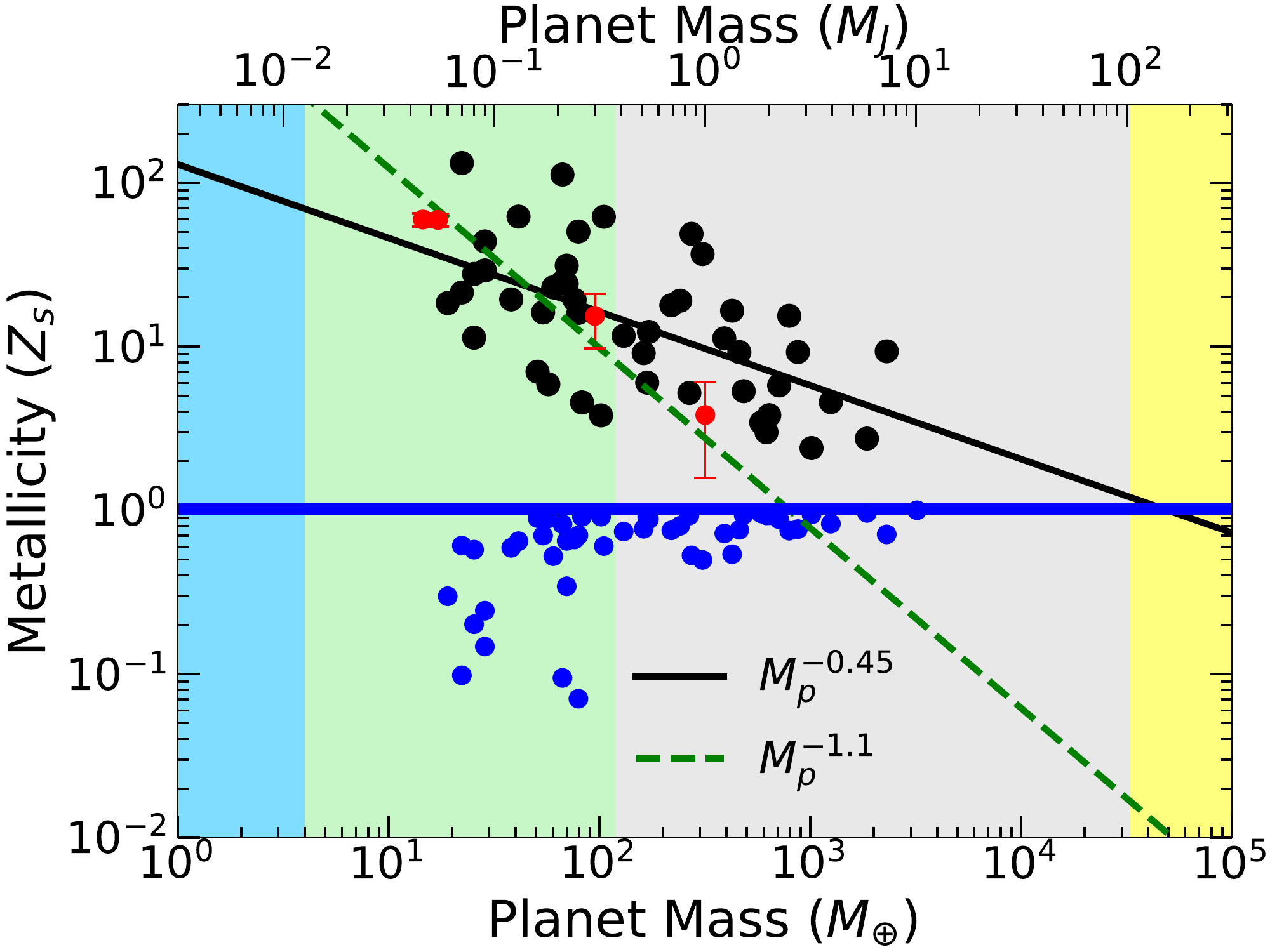}
\caption{Characterization of observed exoplanets based on our analyses (also see Table \ref{table4}). 
On both panels, the computed values and the best fit of T16 are denoted by the black dots and the black solid line, respectively.
In addition, four planets in the solar system (Jupiter, Saturn, Uranus and Neptune) are shown by the red dots for comparison purpose \citep{sg04,hap11,whm17}.
Note that the error bars of Uranus and Neptune are so small that they are almost invisible in these plots.
On the left panel, the computed $M_{Z,gas}$ and the straight line of $M_Z=M_p$ are plotted by the blue dots and the green dashed line, respectively.
In addition, the upper limit of $M_{Z,gas}(=0.03 M_p)$ is shown by the blue solid line.
The mass range investigated by T16 is divided into two regions (gray and green), following the gas accretion process (see the blue dots and Section \ref{sec:data_2}).
Based on the behavior of $M_{Z,gas}$ and the intersection between $M_Z \propto M_p^{3/5}$ and $M_{Z,gas}=0.03 M_p$, 
the region of gas-rich giant planets is identified (see the grey region).
The region of gas-poor sub-giant planets is determined by the value of $M_{Z,gas}$ and the intersection between $M_Z \propto M_p^{3/5}$ and $M_Z=M_p$ (see the green region).
Since the intersection between  $M_Z \propto M_p^{3/5}$ and $M_Z=M_p$ defines the boundary beyond which planets can contain gaseous atmospheres,
we can suggest that the critical core mass for initiating gas accretion is about $4 M_{\oplus}$ (see the red horizontal line).
In other words, rocky super-Earths will distribute in the blue region.
On the right panel, the computed $M_{Z,gas}/(M_p Z_s)$ and the result of \citet{kbd14} are plotted by the blue dots and the green dashed line, respectively.
Also, the straight line of $Z_p/Z_s=1$ is denoted by the blue solid line for the reference.
Exoplanets in the grey region can be used for studying the metallicity evolution in these planets' atmospheres,
while planets in the green region may suggest a possibility of dissolving planetary cores into their envelopes.
}
\label{fig7}
\end{center}
\end{minipage}
\end{figure*}

In this section, we combine the analyses and discussions done in the above sections.
Keeping the limitations of our analysis in mind (Section \ref{sec:disc_1}),
we develop a comprehensive picture of how observed exoplanets likely formed 
and of how our understanding of planet formation can be improved (see Figure \ref{fig7} and Table \ref{table4}).

We begin with the $M_Z-M_p$ relation (see the left panel of Figure \ref{fig7}). 
We first divide the mass range explored by T16 into two regimes, 
based on the gas accretion process (see the blue dots and the green and grey regions, also see Section \ref{sec:data_2}).
Our analysis suggests that
the behavior of $M_Z$ for planets with the mass of $100M_{\oplus} \la M_p \la  3 \times 10^4 M_{\oplus}$
can be understood well if the observed planets keep the formation histories at their final stages.
It is interesting to point out that this region can be extended to the mass range of brown dwarfs,
beyond which gas accretion plays a more important role in determining the value of $M_Z$ (see the shaded region with the yellow color).
This implies that while their formation efficiency may not be high, some brown dwarfs may form via the same mechanisms of forming planets.
As $M_p$ increases, the contribution ($M_{Z,gas}$) coming from gas accretion becomes more significant (see the blue line).
When $M_p$ reaches $3 \times 10^4 M_{\oplus} (\simeq10^2 M_{J})$, $M_Z \approx  M_{Z,gas}$ (see the yellow region).
In this paper, we tentatively call objects residing in the yellow region as "stars".
Note that the theoretical distinction between a star and a planet should come from formation mechanisms \citep[e.g.,][]{cjj14,hr15}.

We now discuss the lower mass region.
We first mention that the interpretation that is developed for the grey shaded region is still applicable 
to planets that have masses of $15M_{\oplus} \la M_p \la  100 M_{\oplus}$ (see the green region).
In fact, the best fit is obtained for exoplanets with the mass range of $20M_{\oplus} \la M_p \la  3 \times 10^3 M_{\oplus}$ in the original analysis (T16).
It is nonetheless important to emphasize that some planets in the green region did not undergo efficient gas accretion (see the blue dots).
Additional explanations would be needed to fully understand these planets,
which remains to be explored in the future work.

Another interesting point on the left panel of Figure \ref{fig7} is that 
the line of $M_Z \propto M_p^{3/5}$ and the straight line of $M_Z=M_p$ intersect at $M_p \simeq 4 M_{\oplus}$ (see the red solid line).
This indicates that planets with the mass of $> 4 M_{\oplus}$ can essentially obtain gaseous envelopes from their natal protoplanetary disks.
This in turn implies that the critical core mass for the onset of gas accretion is about $4 M_{\oplus}$.
It should be noticed that this value of the core mass is roughly consistent with previous studies
which show that exoplanets with the radius of larger than $\simeq 1.6$ earth radii (the corresponding mass of $\simeq 5-6 M_{\oplus}$) 
are unlikely to be purely rocky \citep[e.g.,][]{mwp14,r15,h16}.
We can therefore suggest that planets less massive than $\simeq 4 M_{\oplus}$ tend to be made mostly of rocky (or solid) materials.
Thus, the $M_Z-M_p$ diagram is  useful for developing a better understanding of planet formation.

Finally, we turn our attention to the $Z_p-M_p$ relation (see the right panel of Figure \ref{fig7}).  
Our analysis suggests that the evolution of atmospheric metallicity in (exo)planets can be explored in the grey region,
by comparing the total planet metallicity ($Z_p$, the black solid line) with the atmospheric metallicity (the green dashed line).
In the entire grey region, gas accretion provides only a minor contribution to $Z_p$ (see the blue dots).
As a result, in order to fully understand the composition of (exo)planet atmospheres and to reliably make a link with planet formation,
a number of physical processes should be taken into account self-consistently.
These are not only gas and solid accretion onto growing planets, 
but also the subsequent processes such as planetesimal ablation in planetary envelopes \citep[e.g,][]{ppr88}, and dust growth and settling there \citep[e.g.,][]{mbp10,mka14,o14}.
For the low mass region, there is not a clear difference between $Z_p$ and the atmospheric metallicity,
which might imply that most masses of planetary cores would potentially dissolve into their envelopes.
Further analysis and/or modeling are certainly required for carefully examining exoplanets in the green region.

In summary, we propose that observed exoplanet populations can be classified into three categories, depending on their masses (see Table \ref{table4}).
When $M_p \la  4 M_{\oplus}$, planets are made predominantly of rocky (or solid) materials, and they can be regarded as (super-)Earths.
When $4M_{\oplus} \la M_p \la  100 M_{\oplus}$,
they contain some amount of gas, so that they can be called as gas-poor, sub-giant planets like Neptune in the solar system.
Note that some mechanisms and/or fine-tuning of the formation timing are necessary for fully reproducing these planets.
This additional requirement leads to a prediction that the population of planets in this category may not be so common \citep[e.g.,][]{il04i,mab09}.
Such a prediction would be effective only for massive ($\ga 10 M_{\oplus}$) planets,
since a couple of formation mechanisms are proposed for mini-Neptune mass planets \citep[e.g.,][]{hm13,cl13}.
Finally, planets that have the mass of $0.4 M_{J} \la M_p \la  10^2 M_{J}$ are viewed as gas-rich giant planets.
It is important to realize that most of their masses originate from their gaseous disks,
while most of heavy elements in these planets are determined by solid accretion such as planetesimals in the last formation stage.

\subsection{Potential roles of the current and future observations} \label{sec:disc_5}

We finally discuss potential roles of the current and future observations of exoplanets and their atmospheres that can be deduced from this work.

We begin with listing up these roles.
Observations of exoplanets' atmospheres will allow one to explore evolution of atmospheric metallicity 
as shown in the right panel of Figure \ref{fig7} (see the grey region).
Especially, comparison between hot and warms Jupiters would be invaluable for investigating how dust growth and settling take place in exoplanetary atmospheres.
This is because the atmospheres of hot Jupiters are considered as fully radiative \citep[e.g.,][]{fnb07},
and atmospheric metallicity may have a steeper slope due to efficient dust growth and settling (see the green dashed line).
On the contrary, warm Jupiters would tend to have convective atmospheres because they are far away from their host stars.
If this would be the case, a larger amount of heavy elements may be able to stay in planetary atmospheres,
and their trend in atmospheric metallicity may differ from that of hot Jupiters.
In addition, observations taken towards brown dwarfs would be interesting for examining their formation origins.
Finally, the dots in the green region are not large enough to fully understand why some planets in this region did not undergo runaway gas accretion.
More observations are obviously needed to investigate how understanding of planetesimal accretion can be extended to this green region \citep[e.g.,][]{mvm16,efm17}
and/or how pebble accretion comes into play to develop better understanding of planet formation \citep[e.g.,][]{mbj17}.

\begin{figure}
\begin{center}
\includegraphics[width=8cm]{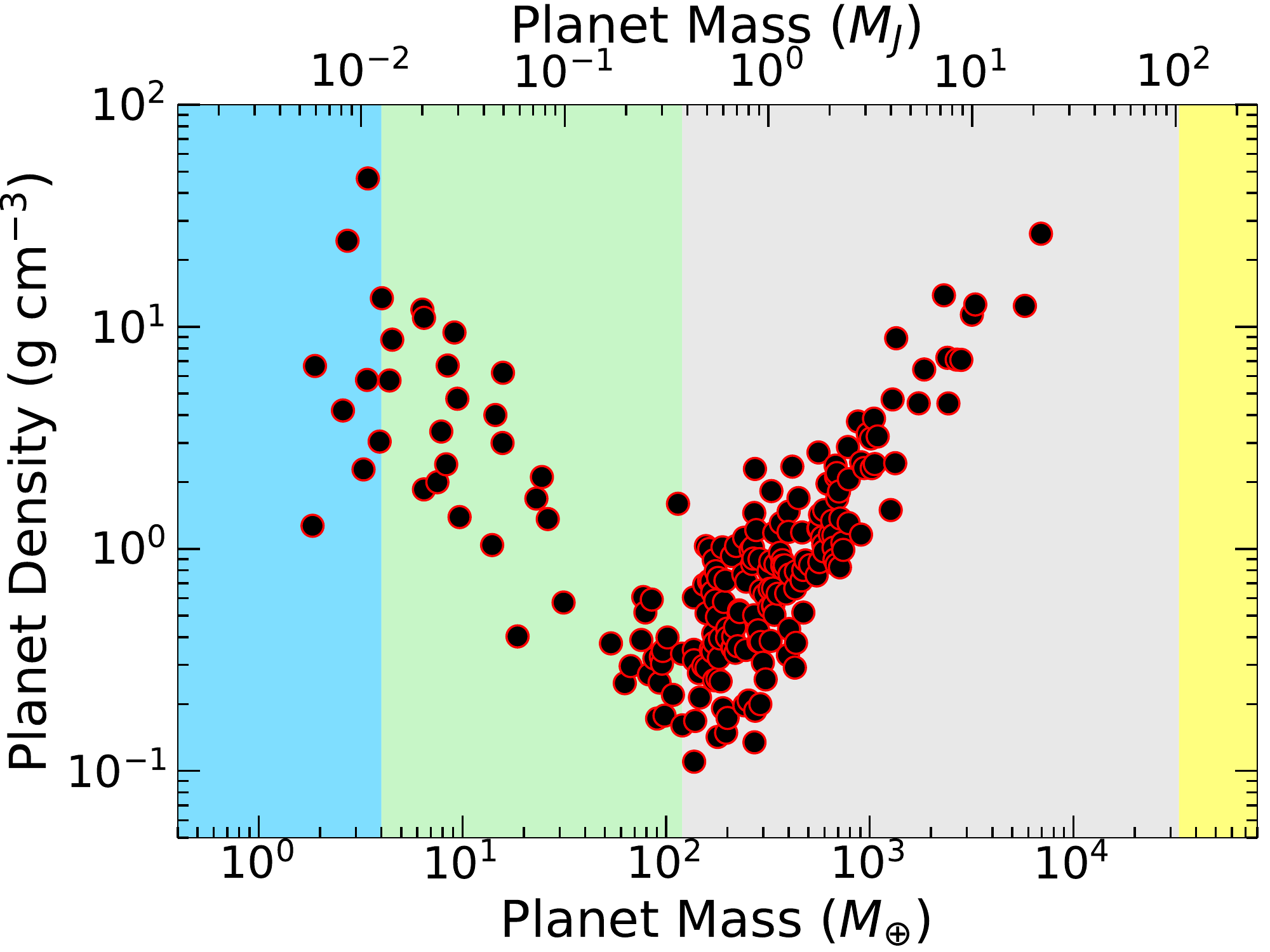}
\caption{Planet density as a function of planet mass. 
As in Figure \ref{fig7}, our classification of planets is shown as the shaded regions (see Table \ref{table4}).
The bulk densities of observed exoplanets are computed directly by adopting the values of planet mass and radius 
taken from the NASA Exoplanet Archive \citep[see the black dots]{acc13}.
It is interesting that the data points in the green region show scatter, which might be related to planetary migration.
Planets in the grey region line up with the straight line with a large band.
This straight line involves the equation of state for metallic hydrogen.
}
\label{fig8}
\end{center}
\end{figure}

How can we examine the effect of planetary migration through observations of exoplanets and their atmosphere?
In order to address this issue, we consider the bulk density of observed exoplanets.
Figure \ref{fig8} shows the data points obtained from the NASA Exoplanet Archive \citep{acc13} with our classification of planets (see Table \ref{table4}).
One interesting feature of this figure is that the data points in the green region have scatter. 
As demonstrated clearly in Figure \ref{fig3}, planets in this region are most sensitive to subtraction of the assumed core mass.
This in turn suggests that the total heavy element mass is regulated predominantly by the core mass itself.
Given that the contribution of planets' atmospheres is not so significant to their total mass (see Figure \ref{fig2}),
this scatter may be interpreted as a potential signature of planetary migration: 
when core formation takes place in various regions of protoplanetary disks, 
their bulk densities posses diversity.
If the subsequent planetary migration delivers these cores in the current positions,
the observed diversity can be used as a fossil record of where they form in the disks.
Another interpretation of Figure \ref{fig8} is that while the contribution of H/He-dominated atmospheres is not substantial to the total mass,
their contribution to the planet radius is crucial, leading to diversity in the bulk density of planets \citep[e.g.,][]{wl15}.

\section{Summary \& Conclusions} \label{sec:conc}

We have investigated how accretion of gas and solids onto growing planets determines the trend of the total heavy element mass ($M_Z$) in observed exoplanets.
This work is motivated by T16 which shows that 
observed exoplanets have the correlations of $M_Z \propto M_p^{3/5}$ and $Z_p/Z_s\propto M_p^{-2/5}$ (see Figure \ref{fig1} and Table \ref{table3}).

We have made use of the existing semi-analytical formulae that are derived from more detailed studies,
and explored how accretion of planetesimals and pebbles proceeds onto planets with simultaneous gas accretion (see Table \ref{table3}).
We have demonstrated that the $M_{Z}-M_p$ relation discovered by T16 is understood well if the relation traces the final stage of planet formation. 
At the stage, planets accrete solids from their gapped planetesimal disks and gas accretion is limited by disk evolution.
We have also found that core formation and pebble accretion cannot reproduce the power-law index derived by T16.
It is interesting that this work suggests that pebble accretion might not play a direct role at the final formation stage.
Moreover, our analysis has showed that the contribution arising from gas accretion is negligible to the total heavy element mass in planets (see Figure \ref{fig5}).

We have then reanalyzed the results of T16 to consider how they can be used for deriving some insights about planet formation.
We have found that the envelope mass becomes comparable to the total planet mass at $M_p > 100 M_{\oplus}$ (see Figure \ref{fig2}).
It is interesting that some planets in the mass range of $20 M_{\oplus} \la M_p \la 100 M_{\oplus}$ have less massive envelopes, compared with the total mass.
This indicates that they did not undergo runaway gas accretion.
Some mechanisms and/or fine tuning of formation timing are needed for postponing the onset of runaway gas accretion for these planets.
Furthermore, we have applied the results of our analysis to the atmospheric metallicity of exoplanets.
Our analysis has suggested that the evolution of metallicity in exoplanets' atmospheres can be examined 
by comparing the total heavy element mass in planets and the heavy element mass in their atmospheres (see Figure \ref{fig6}). 

We have compared our results with those of previous studies (see Table \ref{table3}).
We have found that despite of the simplicity of our analysis, we can reproduce the power-law index of the $Z_p-M_p$ relation that is obtained by \citet{mka14},
when the same assumption is employed.
Note that the power-law index of \citet{mka14} is different from that of T16.
We can therefore conclude that 
our simplified, but physically motivated framework provides a clearer view of 
under what conditions the correlations of $M_Z \propto M_p^{3/5}$ and $Z_p \propto M_p^{-2/5}$ are generated in the course of planet formation.

We have listed up the limitations of our analysis that should be examined by detailed numerical simulations and the future observations.
We have discussed other physical processes that are not included in our analysis, such as planetary migration and the effect of multiple planet formation.
Combining all the analyses done in this paper, we have proposed a classification of observed exoplanets.
We have finally summarized potential roles of the current and future observations of exoplanets and their atmospheres.
It is important to detect exoplanets' atmospheres more for exploring the evolution of metallicity there.

Thus, we conclude that investigation of the the $M_{Z}-M_p$ relation is very important for understanding the final stage of planet formation.
And further detailed modeling, numerical simulations, 
and dealing with a larger number of observational data are required for confirming our results and drawing a more complete picture of planet formation.


\acknowledgments

The authors thank an anonymous referee for useful comments on our manuscript.
Y.H. thanks Jonathan Fortney for his encouragement of this work and Gennaro D{'}Angelo for stimulating discussions.
This research was carried out at the Jet Propulsion Laboratory, California Institute of Technology,
under a contract with the National Aeronautics and Space Administration.
Y.H. is supported by JPL/Caltech.

\appendix

\section{Effect of gas gaps on the $M_{Z}-M_p$ relation} \label{app1}

Here we briefly discuss how the presence of gas gaps will affect the $M_{Z}-M_p$ relation.

To proceed, we make use of the results obtained by \citet{ti07}.
In this study, mass growth of planets via gas accretion is investigated.
In order to reliably take into account the effect of gas gaps that are opened up by planets,
they consider both tidal interaction arising from the planets and viscous diffusion of gas disks,
and compute the resulting value of $\sigma_{p,acc}$ that regulates gas accretion onto the planets (see equation (\ref{eq:tau_hydro})).
They find that the profile of $\sigma_{p,acc}$ can divide into two regions.
Provided that $M_s$, $r_p$, $\Omega_p$, $h$, and $\nu$ are the stellar mass, the position of a planet, the angular frequency at $r=r_p$, 
the disk pressure scale height, and the disk viscosity, respectively,
the the Hill radius of the planet ($r_H$), the characteristic lengths, $l$ and $x_m$, are given as
\begin{equation}
r_H =  \left( \frac{ M_p }{ 3 M_s } \right)^{1/3} r_p \propto M_p^{1/3},
\end{equation}
\begin{equation}
l =  \left[ \frac{8}{81 \pi}  \left( \frac{ \nu }{ r_p^2 \Omega_p } \right)^{-1}  \left( \frac{ M_p }{ M_s} \right)^{2} \right]^{1/3} r_p \propto M_p^{2/3},
\end{equation}
and
\begin{equation}
x_m =  12^{1/5}   \left( \frac{ h }{ l } \right)^{2/5} l.
\end{equation}
Then, the profile of $\sigma_{p,acc}$ is determined by the balance between tidal torque and viscous diffusion for the case of $2 r_H > x_m$.
For the case of $2 r_H < x_m$, its profile becomes steep enough that the Rayleigh instability condition eventually regulates the behavior of $\sigma_{p,acc}$.
Finally, the gas accretion timescales can be given as (see equation (B3) in \citet{ti07})
\begin{equation}
\tau_{g,acc}^{Gap} \propto M_p^{1} \mbox{ for } 2r_H > x_m,
\end{equation}
and as (see equation (B8) in \citet{ti07})
\begin{equation}
\tau_{g,acc}^{Gap} \propto - \frac{1}{3} \left( \frac{2 r_H}{h} \right)^2 + \frac{11}{12^{4/5} }  \left( \frac{2 r_H}{h} \right) \left( \frac{l}{h} \right)^{3/5} 
                                           -  \frac{8}{12^{3/5} }  \left( \frac{l}{h} \right)^{6/5}   \mbox{ for } 2r_H < x_m.
\end{equation}

Combining these timescales with the solid accretion rate ($dM_{Z,solid}/dt$), we obtain tthe $M_{Z}-M_p$ relation
(see equations (\ref{eq:dz_dm_pl_nogap}) and (\ref{eq:dz_dm_pl_gap})).
Table \ref{tableA1} summarizes the results.
One may wonder that the trend found by T16 can be reproduced if no gap is formed in planetesimal disks for both the cases of $2r_H > x_m$ and $2r_H < x_m$.
We argue that if gas gaps are already opened up by planets,
then it can be anticipated that gap formation would take place in planetesimal disks as well.
This is because under the presence of gas gaps, planetesimals can obtain a higher value of eccentricity there,
which arises from both the high mass of planets and a reduced efficiency of eccentricity damping by the disk gas.
Once such planetesimals enter the gas rich region that is beyond gas gaps, 
however, their eccentricity can rapidly decrease, and hence gap formation in planetesimal disks can be accelerated.

Thus, when gaps are present around planets in gas disks, it would be reasonable to consider that planetesimal disks also have gaps.
Under such a condition, the $M_Z-M_p$ relation that is derived from observed exoplanets cannot be reproduced.

\begin{table*}
\begin{minipage}{17cm}
\begin{center}
\caption{Power-law indices for the $M_Z-M_p$ relation when the effect of gas gaps is taken into account}
\label{tableA1}
\begin{tabular}{c|c|c|c|c} 
\hline 
                                                 & $2r_H > x_m$   & $2r_H < x_m$             & $2r_H < x_m$                                   & $2r_H < x_m$                   \\  
                                                 &($D=1$)              &  First term ($D=2/3$)  &  Second term ($D=11/15$)                &  Third term  ($D=4/5$)      \\   \hline   \hline                                           
No gap in planetesimal disks   &  3/5                    & 8/15 $\simeq$ 1/2       &  41/75 $\simeq$ 8/15 $\simeq$ 1/2   &  14/25 $\simeq$ 3/5          \\  
gap in planetesimal disks         &  1/30                  & 1/6                               &  7/50 $\simeq$ 1/7                            &   17/150 $\simeq$ 3/25 $\simeq$ 1/8  \\ 
\hline 
\end{tabular}
\end{center}
\end{minipage}
\end{table*}






\bibliographystyle{apj}          

\bibliography{apj-jour,adsbibliography}    

\end{document}